\PassOptionsToPackage{backref=page}{hyperref}

\documentclass[fleqn,usenatbib]{mnras}

\usepackage[T1]{fontenc}

\DeclareRobustCommand{\VAN}[3]{#2}
\let\VANthebibliography\thebibliography
\def\thebibliography{\DeclareRobustCommand{\VAN}[3]{##3}\VANthebibliography}

\usepackage{graphicx}	
\usepackage{amsmath}	
\usepackage{amssymb}	
\usepackage[export]{adjustbox}
\usepackage{caption}
\usepackage{subcaption}
\usepackage{makecell}
\usepackage{comment}
\usepackage{enumitem}
\usepackage{float}
\usepackage{hhline}
\usepackage{newtxtext,newtxmath}

\title[Environmental effects on galaxy colour]{Red riding on hood: Exploring how galaxy colour depends on environment}

\author[P. C. Bhambhani et al.]{
Pankaj C. Bhambhani,$^{1}$\thanks{E-mail: pcb.astro@gmail.com}
Ivan. K. Baldry,$^{1}$\thanks{E-mail:
I.Baldry@ljmu.ac.uk}
Sarah Brough,$^{2}$
Alexander D. Hill,$^{1,6}$
\newauthor\,M. A. Lara-Lopez,$^{3}$
J. Loveday,$^{4}$
B.W. Holwerda,$^{5}$
\\\\
$^{1}$Astrophysics Research Institute, Liverpool John Moores University, IC2, Liverpool Science Park, 146 Brownlow Hill, Liverpool L3 5RF, UK\\
$^{2}$School of Physics, University of New South Wales, NSW 2052, Australia\\
$^{3}$Departamento de F\'{ı}sica de la Tierra y Astrof\'{ı}sica, Instituto de F\'{ı}sica de Part\'{ı}culas y del Cosmos, IPARCOS. \\Universidad Complutense de Madrid (UCM), E-28040, Madrid, Spain\\
$^{4}$Astronomy Centre, University of Sussex, Falmer, Brighton BN1 9QH, UK\\
$^{5}$Department of Physics and Astronomy, 102 Natural Science Building, University of Louisville, Louisville KY 40292, USA\\
$^{6}$Department of Physics, University of Liverpool, Liverpool L69 7ZE, UK
}

\date{Draft version July 2022}

\pubyear{2022}

\begin{document}
\label{firstpage}
\pagerange{\pageref{firstpage}--\pageref{lastpage}}
\maketitle

\begin{abstract}
Galaxy populations are known to exhibit a strong colour bimodality, corresponding to blue star-forming and red quiescent subpopulations. The relative abundance of the two populations has been found to vary with stellar mass and environment.
In this paper, we explore the effect of environment considering different types of measurements. We choose a sample of \(49,911\) galaxies with \(0.05 < z < 0.18\) from the Galaxy And Mass Assembly survey. We study the dependence of the fraction of red galaxies on different measures of the local environment as well as the large-scale "geometric" environment defined by density gradients in the surrounding cosmic web.
We find that the red galaxy fraction varies with the environment at fixed stellar mass. The red fraction depends more strongly on local environmental measures than on large-scale geometric environment measures. By comparing the different environmental densities, we show that no density measurement fully explains the observed environmental red fraction variation, suggesting the different measures of environmental density contain different information. We test whether the local environmental measures, when combined together, can explain all the observed environmental red fraction variation. The geometric environment has a small residual effect, and this effect is larger for voids than any other type of geometric environment. 
This could provide a test of the physics applied to cosmological-scale galaxy evolution simulations
as it combines large-scale effects with local environmental impact.
\end{abstract}





\section{Introduction}

The field of galaxy formation and evolution tackles the task of explaining how a universe with a spatial uniformity of 1 in \(10^5\) - as evident from the Cosmic Microwave Background (CMB) radiation and other observations - evolved into the complex population of galaxies we see today. This population exhibits strong variations in a diverse set of properties including stellar mass, luminosity, size, morphology, colour, star formation rate (SFR), mean stellar age, etc. (e.g. \citealt{straussScienceBreakthroughsSloan2002,peacockImplications2dFGRS2003,driverGAMAPhysicalUnderstanding2009})

The framework currently favoured by cosmologists for explaining these properties is the \(\Lambda\)CDM model. Here cold dark matter (CDM) haloes form hierarchically from local overdensities, and the baryons fall into the gravitational potential wells of these halos and form galaxies (e.g. \citealt{1997ApJ...490..493N}). However, there are many details not well understood, especially due to the complex interactions of baryons with other baryons and with dark matter. 

A wide variety of models attempting to produce realistic galaxy populations exist. These may be broadly separated into two classes: semi-analytic models (SAMs) and hydrodynamical simulations. SAMs predict the formation and evolution of galaxies by combining the merger trees of dark matter-only N-body simulations with analytic prescriptions of baryonic physics (e.g. \citealt{bowerBreakingHierarchyGalaxy2006, crotonManyLivesActive2006,2008MNRAS.391..481S,lagosPredictionsCOEmission2012,2017MNRAS.465...32B}).
Hydrodynamical simulations, in contrast, evolve dark matter and baryons simultaneously and broadly self-consistently  \citep{schaye10, dubois14, hopkins14, vogelsberger14, schaye15, crain15, springel18, dave19}. Baryonic processes taking place under the resolution limit of the simulation 
are modelled via analytic ‘subgrid’ routines.

Both classes of model are naturally informed by observations of galaxy populations. Recent large scale multi-wavelength galaxy surveys provide a wealth of observations which enable an improved understanding of galaxy formation and evolution.

\subsection{Galaxy Colour Bimodality}
\label{sec:intro_galaxy_bimodality}


The advent of large scale galaxy surveys have helped discover a range of features in galaxy properties. One such interesting feature is the bimodality in galaxy colour. \cite{stratevaColorSeparationGalaxy2001} first observed this in the observed-frame colour using data from the Sloan Digital Sky Survey (SDSS, \citealt{stoughtonSloanDigitalSky2002}). \cite{baldryQuantifyingBimodalColormagnitude2004} analyzed the effect in the rest-frame colour when looking at the colour magnitude relations using SDSS data.

The colour-magnitude diagram (CMD) is an important diagnostic relation in astronomy, particularly in galaxy evolution. The absolute magnitude of a galaxy is its integrated starlight and is a proxy for stellar mass. The colour encodes information on the ages of the contributing stellar populations, i.e. the star formation history -- barring extinction caused by dust. A CMD thus helps understand the evolution of star formation in the galaxy as a function of its stellar mass. Recent works (e.g. \citealt{taylorGalaxyMassAssembly2015}) have used stellar population synthesis (SPS, \citealt{taylorGalaxyMassAssembly2011}) to estimate the stellar mass, and have used colour-mass diagrams instead of colour-magnitude diagrams.

Subsequent research into colour bimodality has validated the findings of \cite{stratevaColorSeparationGalaxy2001} and \cite{baldryColorBimodalityImplications2004} to at least \(z < 2\) (e.g. \citealt{2004ApJ...615L.101B,2004ApJ...600L..11B,2009ApJ...691.1879W,taylorGalaxyMassAssembly2015,2018ApJ...866..136F}). The colour bimodality is interesting to theorists as well because colour has a direct relation with star formation history.

When looking at just the colour of galaxies, the corresponding distribution appears to be well approximated as a sum of two Gaussian distributions (e.g. \citealt{stratevaColorSeparationGalaxy2001,taylorGalaxyMassAssembly2015}), 
This suggests that galaxy formation processes given rise to two different dominant galaxy populations, each with a different distribution of colour values. The two populations are generally labeled "blue" and "red" based on their lower and higher colour values respectively, though their exact definitions have varied from one research work to another (e.g. \citealt{taylorGalaxyMassAssembly2015}). Studies of this bimodality at different redshifts have revealed another interesting finding - although the populations are roughly equivalent in total stellar mass at \(z \sim 1\), the red population has nearly doubled in stellar mass, stellar mass density, and number density over the past \(\sim 7\) Gyr (e.g \citealt{2004ApJ...600L..11B,2007A&A...476..137A,2018ApJ...866..136F}). This suggests that galaxies move between the two populations - these are two evolutionary stages. But the existence of two distinct colour modes must mean the transition from blue to red stage is relatively quick (e.g. \citealt{2004ApJ...615L.101B}). 

In order to reproduce the observed colour distribution and its evolution with redshift, theoretical models tend to add mechanisms that rapidly disrupt or prevent star formation in galaxies. This process is generically called "quenching" (see e.g. \citealt{2010ApJ...721..193P}), and it mainly involves removal of star forming gases. 
While the exact nature of quenching remains debated, popular candidates involve feedback from an Active Galactic Nuclei (AGN) (e.g. \citealt{bensonWhatShapesLuminosity2003,bowerBreakingHierarchyGalaxy2006,crotonManyLivesActive2006,2008MNRAS.391..481S}).

\subsection{The Effect of Environment}
\label{sec:intro_env_effects}



In addition to stellar mass, galaxy colour is also found to be dependent on the local environment. Previous works have identified that galaxies are likely to be red in denser environments (e.g. \citealt{2004MNRAS.353..713K,baldryGalaxyBimodalityStellar2006,2018A&A...618A.140V,2021MNRAS.506.3364R}). Previous studies have found an increase in the local environment density to correlate with a decrease in the star-formation rate \citep{schaeferSAMIGalaxySurvey2017a,schaeferSAMIGalaxySurvey2019}, a decrease in the number of star-forming galaxies \citep{barsantiGalaxyMassAssembly2018} or a change in the stellar kinematic properties of galaxies \citep{vandesandeSAMIGalaxySurvey2021}. The most widely agreed reason for environmental dependence is therefore the loss of star forming gas as a result of stripping in denser environments (e.g. \citealt{barsantiGalaxyMassAssembly2018,trusslerBothStarvationOutflows2020,sotillo-ramosGalaxyMassAssembly2021}). The major processes involved include galaxy harassment and mergers(e.g. \citealt{bialasOccurrenceGalaxyHarassment2015}), strangulation (e.g. \citealt{pengStrangulationPrimaryMechanism2015} and ram pressure stripping\citep{1972ApJ...176....1G,broughGalaxyMassAssembly2013}.

The environment of a galaxy, broadly speaking, is the region surrounding it that has a potential to interact with the galaxy. The term environmental density (also called an environmental measure) refers to a metric that characterises the kind of environment a galaxy is located in.
Just like for the terms red and blue (as mentioned above), there are many commonly used definitions for environment. These include the number of (other) galaxies in a given region (e.g. \citealt{baldryGalaxyBimodalityStellar2006,yoonSpectrophotometricSearchGalaxy2008,2010ApJ...721..193P,broughGalaxyMassAssembly2013,liskeGalaxyMassAssembly2015,schaeferSAMIGalaxySurvey2017a, baldryGalaxyMassAssembly2018,vandesandeSAMIGalaxySurvey2021}), the type of large scale structure surrounding it (e.g. a void, a knot, etc, see \citealt{alpaslanGalaxyMassAssembly2014, eardleyGalaxyMassAssembly2015}),  group/cluster virial mass and distance from the cluster centre (e.g. \citealt{barsantiGalaxyMassAssembly2018, 2018A&A...618A.140V} ), position as a central/satellite galaxy in a halo (e.g. \citealt{2016MNRAS.456.4364B}), and identifying galaxy groups using friends-of-friends based grouping algorithm \citep{robothamGalaxyMassAssembly2011,schaeferSAMIGalaxySurvey2019}. The source of environmental dependence is generally thought to be the local overdensity (e.g. \citealt{baldryGalaxyBimodalityStellar2006, behrooziUniverseMachineCorrelationGalaxy2019}) but there may also be a possible effect of the region of the cosmic web in which the galaxy lies (e.g. \citealt{eardleyGalaxyMassAssembly2015}). The former type of dependence is termed the local environment, while the latter is often termed the large-scale environment. In accordance with \cite{eardleyGalaxyMassAssembly2015}, we use the term use the term `geometric environment' to denote the different regions of the cosmic web.

\cite{baldryGalaxyBimodalityStellar2006} investigated the effects of local environment on galaxy colour in detail. In particular, they studied how the double Gaussian fits to the colour distribution changed with the projected density of neighbouring galaxies. They found that increasing the environmental density has two effects - a major effect is the transition of galaxies from the blue Gaussian distribution to the red one, and a minor effect is the modest reddening of each Gaussian fit, i.e. the mean shifting to slightly higher colour values. They also found that within each environmental density bin, the data were always well modeled by a double Gaussian fit. This suggests that the galaxy properties within the blue and red distributions do not change much with environment.

In contrast, \cite{alpaslanGalaxyMassAssembly2015} investigated the effect of galaxy colour (and other properties) as a function of non-local environment. One of the environmental measures used by them was 
derived from classifications of the large scale structure, 
as obtained from the GAMA Large Scale Structure Catalogue \citep{alpaslanGalaxyMassAssembly2014}.
Their results do not show a significant difference in galaxy colour (or any other property) when comparing across the different large scale structure environments.

There has also been work done that compares the effect of different environments. \cite{perezGlobalEnvironmentalEffects2009} study the effects of galaxy interactions and mergers in different local density and host-halo mass environments, using data from SDSS DR4. \cite{2005MNRAS.358...88W} obtain independent measurements of density on different scales and use that to study galaxy colour dependence in SDSS DR5 data. \cite{pandeyExploringGalaxyColour2020} use spectroscopic data from SDSS DR16 to explore galaxy colour in different environments of the cosmic web.


Environmental effects in hydrodynamical simulations arise naturally from the evolution in the properties of individual gas particles. The change in these properties are computed through a combination of hydrodynamical schemes such as SPH \citep{hopkins13}, and subgrid processes, e.g. star formation and radiative cooling \citep{wiersma09}. The influence of environmental processes such as ram-pressure stripping in quenching star-formation within simulated galaxies is a popular area of study \citep[e.g.][]{trayford16, simpson18}. Discrepancy between simulations in the influence of environment arise due to differences in resolution, subgrid treatment, and the hydrodynamic solvers employed. SAMs, in contrast, require the inclusion of explicit analytic forms of environmental effects, such 
as galaxy mergers, tidal interactions and ram pressure stripping \citep{tecce10, ayromlou19}. The processes included and the methods employed vary between SAMs.
The success of hydrodynamic simulations and SAMs are largely measured through their ability to match observations. With the advent of large-scale surveys like the SDSS and the Galaxy and Mass Assembly (GAMA) survey (Driver et al. 2009, 2011; Liske et al. 2015; Baldry et al. 2018), we have opportunities to study the environmental dependence in more detail and thereby further test the accuracy of simulations of galaxy formation and evolution.

There is also the potential to study the effect of geometric environment. 
The cosmic web \citep{bond96} has been theorised to affect the properties of galaxies and their host dark matter haloes, with $N$-body simulations and simulations of galaxy formation and evolution playing a key role in this field of study \citep[e.g.][]{hahn2007, wang11, goh19, kraljic2020, hellwing21}. Simulations have also been used to explore the relationship between environment and galaxy properties \citep[e.g.][]{bhowmick20, kristensen21}, as well as the variation in this relationship imparted by different measures of environment \citep[][]{haas12, muldrew12}. It is suggested that measures of environment defined on varied spatial scales are encoded with different information, affected as they are by physical processes to a greater or lesser extent, which then leads to varied correlations with galaxy properties.

\subsection{Aim of this work}

In this paper, we examine the effect of environment and stellar mass on the galaxy colour distribution. We use as a metric the fraction of galaxies within a given mass/environment region that are red. We aim to answer the following questions:
\begin{itemize}
    \item How does the red fraction change as a function of stellar mass and environment?
    \item Do the different environmental measures convey different information?
    \item In particular, can the local environment fully explain the red fraction variation, or is there a residual effect from the geometric environment (i.e. the cosmic web)? 
\end{itemize}

In this sense, our high-level goals are similar to that of  \cite{baldryGalaxyBimodalityStellar2006,2010ApJ...721..193P,alpaslanGalaxyMassAssembly2015} and \cite{taylorGalaxyMassAssembly2015}. We build upon their work, by using the latest data and experimenting with different 
density measures. 
The remainder of this paper is organised as follows: section \ref{data} describes the samples of galaxies used in the investigation, which is taken from the GAMA survey. 
Section \ref{methodology} details the methodology used to compute the
red fraction estimates.
In section \ref{results_discussion}, we present and discuss our findings. 
Section \ref{future_work_conclusion} provides a summary of this work and suggests future research directions. 

Throughout this work, we assume a flat \(\Lambda\)CDM cosmology with the parameters \(H_0 = 70 \textsf{ km/s/Mpc}, \Omega_{m,0} = 0.3, \Omega_{\Lambda,0} = 0.699\). All stellar mass estimates obtained from GAMA assume a \cite{2003PASP..115..763C} initial mass function (IMF). All magnitudes provided by GAMA are given in the AB magnitude system. 
Logarithms are assumed to be with base 10 unless specified otherwise, and abbreviated as \(\log\).

\section{Data} 
\label{data}

\begin{figure*}
	\centering
		\begin{subfigure}[t]{.495\textwidth}
			\centering 
		    
		\includegraphics[width=\textwidth,height=0.25\textheight]{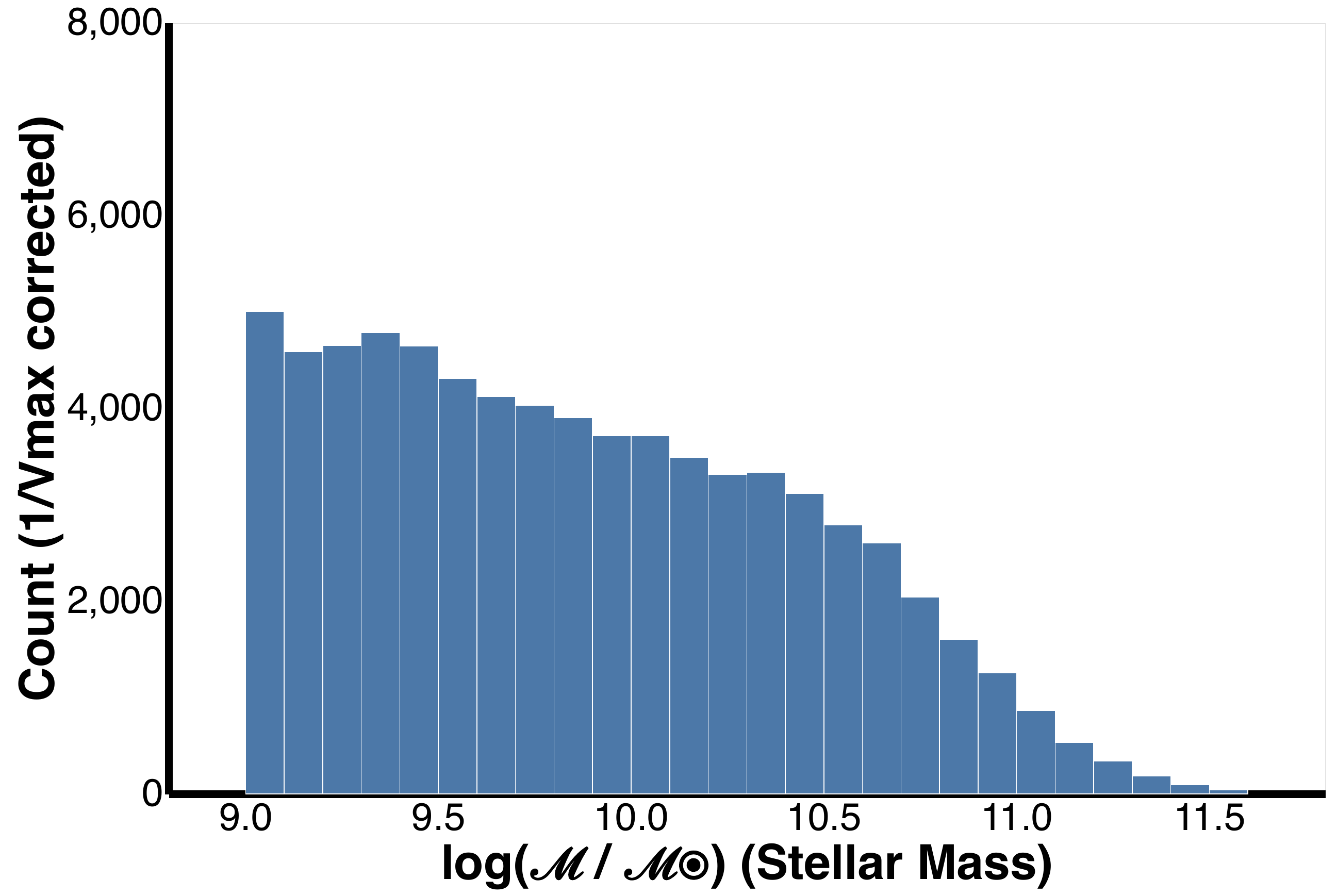}
			\caption
			{}
			\label{fig:data_hist_logmstar}
		\end{subfigure}
		\hfill
		\begin{subfigure}[t]{0.495\textwidth}   
			\centering 
		    
		\includegraphics[width=\textwidth,height=0.25\textheight]{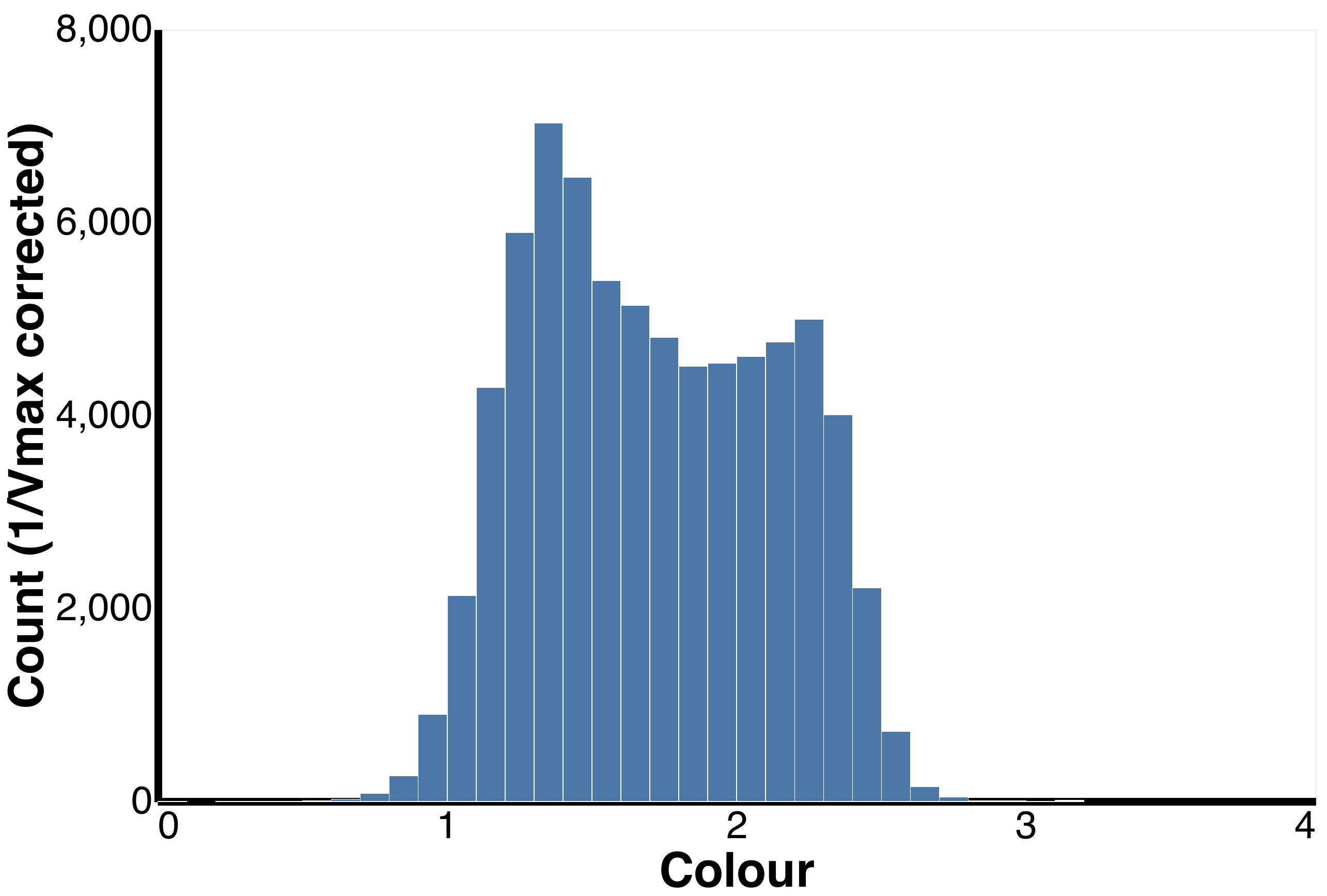}
		\caption
			{}	\label{fig:data_hist_uminusr}
		\end{subfigure}

	\caption
	{{ \small Histograms of (a) mass and (b) colour attributes for the objects in our chosen sample. The counts shown here are volume (\(1/V_{max}\)) weighted, see section \ref{ssec:methodology_vmax} for more information on how this correction is done. }}
	\label{fig:data_mass_colour_histograms}
\end{figure*}

\begin{figure*}
	\centering
		\begin{subfigure}[b]{0.475\textwidth}   
			\centering 
			\includegraphics[width=\textwidth]{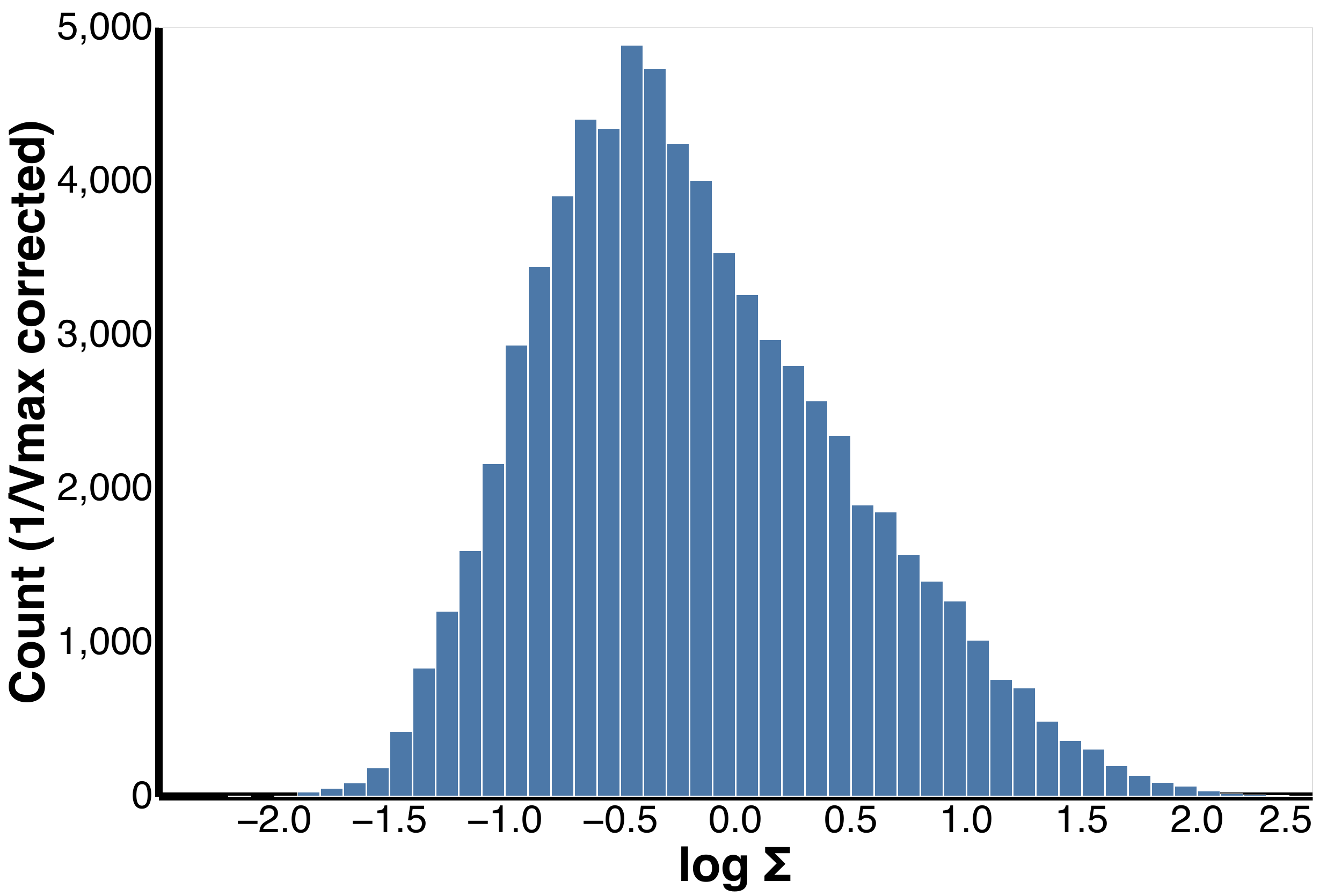}
			\caption[]%
			{{\small  }}    
			\label{fig:data_hist_log_sigma}
		\end{subfigure}
		\hfill
		\begin{subfigure}[b]{0.475\textwidth}   
			\centering 
			\includegraphics[width=\textwidth]{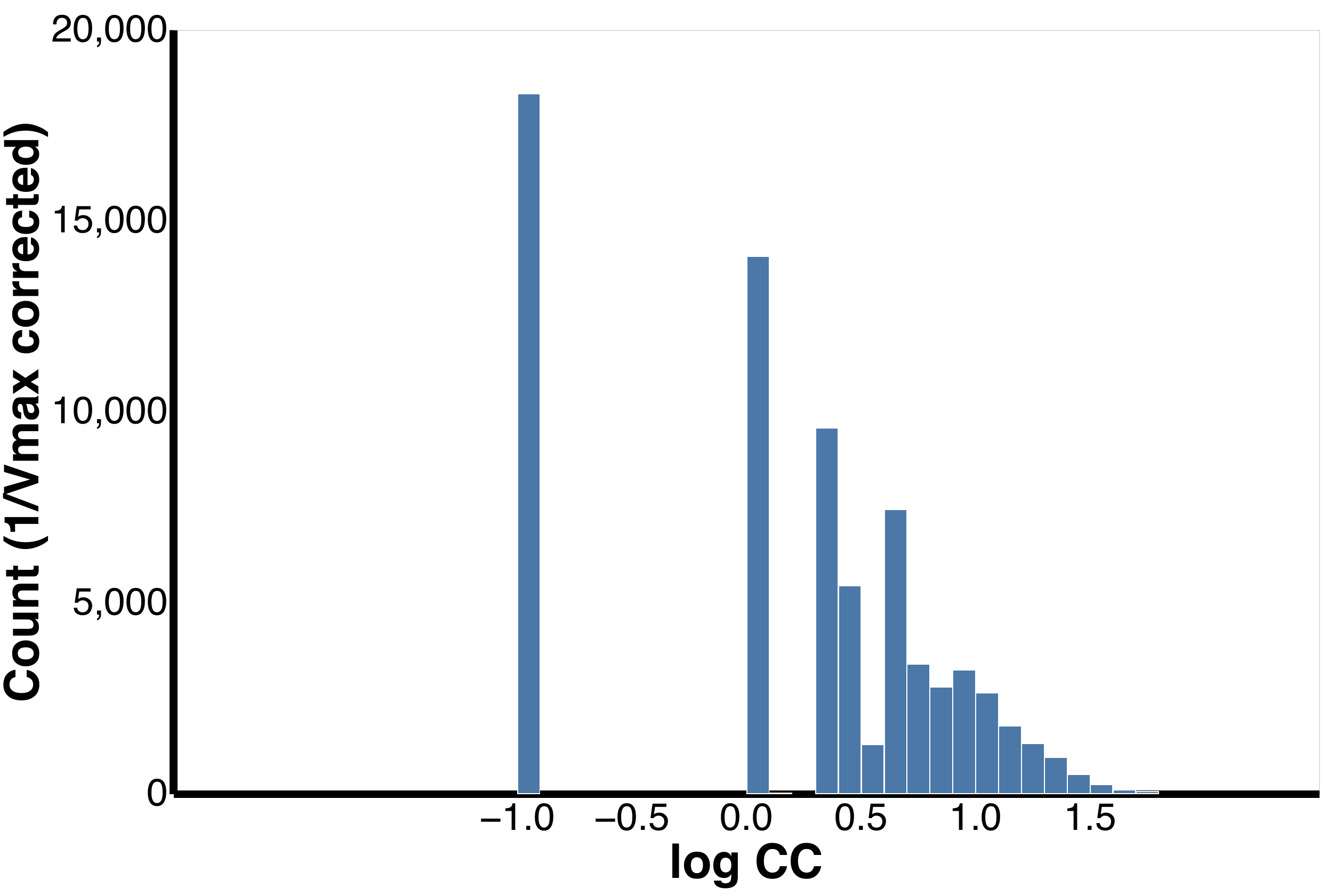}
			\caption[]%
			{{\small }}    
			\label{fig:data_hist_log_cc}
		\end{subfigure}
	\vskip\baselineskip
		\begin{subfigure}[b]{0.475\textwidth}   
			\centering 
			\includegraphics[width=\textwidth]{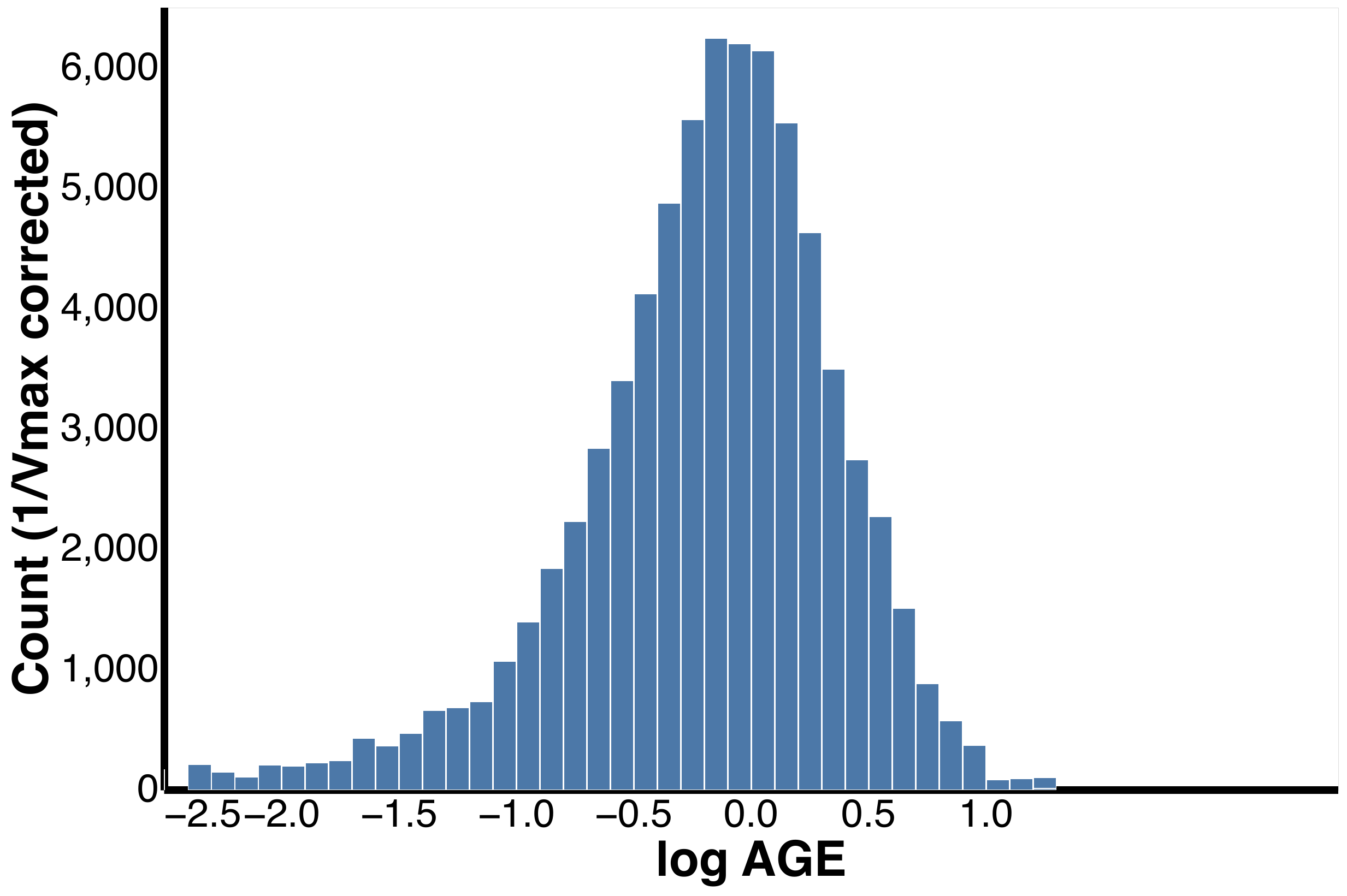}
			\caption[]%
			{{\small }}    
			\label{fig:data_hist_log_age}
		\end{subfigure}
		\hfill
		\begin{subfigure}[b]{0.475\textwidth}   
			\centering 
			\includegraphics[width=\textwidth]{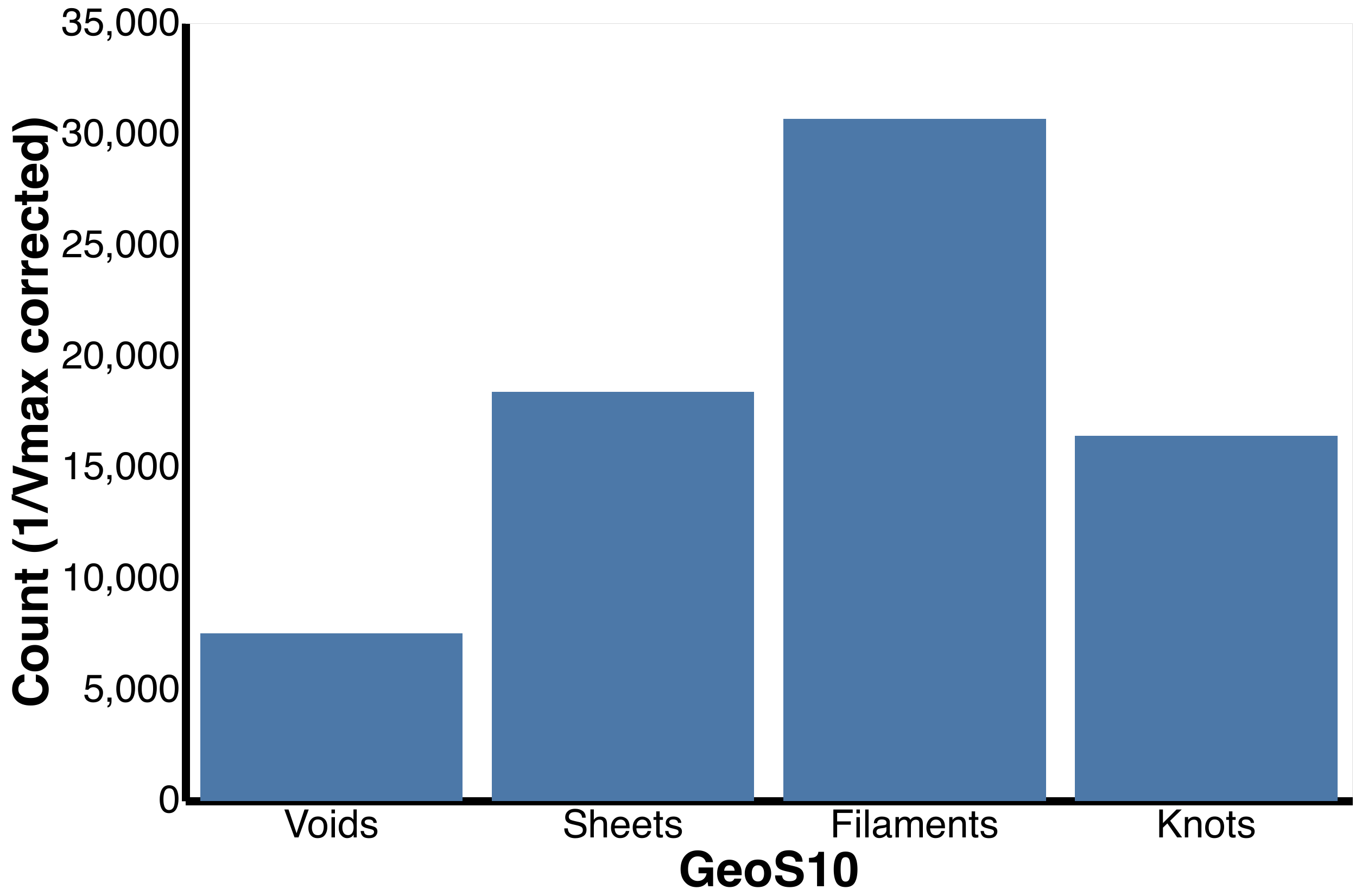}
			\caption[]%
			{{\small }}    
			\label{fig:data_hist_geos10}
		\end{subfigure}
	\caption[ ]
	{{ \small Histograms of the different environmental density measures for the objects in our chosen sample, namely (a) 5th Nearest Neighbour (\(\Sigma\)), (b) Cylindrical Count (CC), (c) Adaptive Gaussian Ellipsoid (AGE) and (d) Geometric Environment with count smoothing scale \(\sigma = 10 h^{-1}\) Mpc (GeoS10). See section \ref{sec:data_env_densities} for details on these environmental measures. The counts shown here are volume (\(1/V_{max}\)) weighted, see section \ref{ssec:methodology_vmax} for more information on how this correction is done. }}
	\label{fig:data_env_density_histograms}
\end{figure*}

			
The Galaxy and Mass Assembly (GAMA) survey \citep{driverGAMAPhysicalUnderstanding2009,2011MNRAS.413..971D,liskeGalaxyMassAssembly2015,baldryGalaxyMassAssembly2018,driverGalaxyMassAssembly2022} is an international large-scale galaxy redshift survey. Its goal is to probe the universe on intermediate scales between 1 kpc and 1 Mpc. While cosmological surveys have improved our understanding of the universe on large scales (> 1 Mpc), the picture is less clear on the scales of galaxies, groups and clusters. The goal of GAMA is to provide observations of structure formation at these scales, including measuring the dark matter halo mass function (HMF), as well as the effect of baryonic feedback through measurements of stellar mass function (SMF) down to very low mass haloes. 
GAMA surveys the sky over 5 regions (three equatorial -- named G09, G12 and G15 and two southern -- named G02 and G23), covering a total area of 286 \(\mathrm{deg}^2\). It provides spectra \citep{hopkinsGalaxyMassAssembly2013} and redshift  \citep{baldryGalaxyMassAssembly2014} estimates from the AAOmega spectrograph as part of the 3.9m Anglo Australian Telescope. It also provides photometric data comprising of optical \textit{ugriz} imaging from SDSS (DR7) and near-infrared (NIR) \textit{ZYJHK} imaging from the VIKING survey. When analyzing environmental effects, it is important to ensure spectroscopic completeness in the sample to avoid selection effects due to local density of galaxies. Through multiple visits to each field \citep{2010PASA...27...76R}, GAMA provides accurate redshift estimates (\(> 98 \%\) confidence) with uniform and near total spectroscopic completeness (\(\gtrsim 98 \%\)) upto a Petrosian \textit{r}-band magnitude limit of \(r_{petro} < 19.8\, \textsf{mag}\), except in G23 which instead has an \textit{i}-band limit of \(19.2 \, \textsf{mag}\). \citep{liskeGalaxyMassAssembly2015,baldryGalaxyMassAssembly2018}.

\subsection{Our Chosen Sample}
Our sample is obtained from
Data Release 4 (DR4, \citealt{driverGalaxyMassAssembly2022}) of the GAMA survey. DR4 has redshift and other information for \(\sim 330,000\) objects. For the local environment, we choose the environment measures defined in the \texttt{EnvironmentMeasures} Data Management Unit (DMU) \citep{broughGalaxyMassAssembly2013,liskeGalaxyMassAssembly2015}. Galaxy environment has been studied previously for GAMA using data products available in the GAMA Galaxy Group Catalogue (G3C, data provided in the \texttt{GroupFinding} DMU), where the neighbours of a galaxies are determined by a Friends of Friends (FoF) linking algorithm. \citep{robothamGalaxyMassAssembly2011}. In a FoF approach, all objects grouped together will have the same number of neighbours. Since we are more interested in environmental measures that vary continuously, we do not include data from G3C in our sample.

We chose limits of \(10^9 \mathcal{M}_\odot \lesssim \mathcal{M} \lesssim 10^{11.6} \mathcal{M}_\odot \) and \(0.05 \lesssim z \lesssim 0.18\), where \(\mathcal{M}\) and \(z\) are the stellar mass and CMB-frame redshift respectively. See sections \ref{ssec:data_stellar_mass} and \ref{ssec:data_redshift} for justification of our choice of limits for mass and redshift respectively. Our sample size is \( 49, 911\) objects, about 16.6\% of all objects in GAMA.

Below we provide a brief description of the different attributes in our samples.



\subsection{Stellar Mass (\(\mathcal{M}/\mathcal{M}_\odot\))}
\label{ssec:data_stellar_mass}

This is the total stellar mass of a galaxy in units of solar mass. 
It is estimated from stellar population synthesis (SPS) models of the measured galaxy spectral energy distributions (SEDs). For more details, see \cite{taylorGalaxyMassAssembly2011} (and section 5.3.2 of \citealt{baldryGalaxyMassAssembly2018}). 
For our sample, we chose a lower limit of \(10^9 \mathcal{M}_\odot\) in line with \cite{baldryGalaxyBimodalityStellar2006}, and an upper limit of  \(10^{11.6} \mathcal{M}_\odot\) due to lack of significant number of higher mass objects. Figure \ref{fig:data_hist_logmstar} shows the stellar mass histogram for objects in our sample.



\subsection{Colour (u - r)} 
\label{sssec:data_colour}
The galaxy colour used here is the difference between the \textit{u} and \textit{r} band magnitudes and is denoted \(u - r\). This colour is obtained from SED fitting \citep{2011MNRAS.418.1587T}, with apertures obtained using LAMBDAR (Lambda Adaptive Multi-Band Deblending Algorithm in R) photometry \citep{2016MNRAS.460..765W}. The values are provided as part of GAMA DR4 and are k-corrected as well as Milky Way dust extinction corrected.
Figure \ref{fig:data_hist_uminusr} shows a histogram of the resulting colour values.

We also experimented with two other versions of colour, derived from the same bands but with variations in photometry and dust extinction correction. We defer to Appendix \ref{ssec:discussion_effect_colour} the discussion of how the different colours affect the red galaxy fraction evolution. We choose the colour \(u - r\) for work as it is derived from more accurate photometry compared to the other two colours and it produces a red fraction dependence consistent with what is known in the literature. 

\subsection{Redshift}
\label{ssec:data_redshift}
There are two redshift attributes in our data set. One is the redshift \(z\) of the object, corrected to the CMB frame. We select \(0.05 \lesssim z \lesssim 0.18\) for our chosen sample. The lower limit is to avoid regions dominated by individual stars. The upper limit is the maximum redshift for which the local environmental measurements are available in GAMA DR4. See section \ref{sec:data_env_densities} for how this redshift upper limit is determined as well for definitions of environmental densities.

The other attribute is the limiting redshift \(z_{max}\) - the maximum redshift at which the  object would be visible for a given \textit{r}-band (Petrosian) magnitude limit. GAMA provides these estimates from iteratively solving the k-correction to a limiting magnitude of \(r < 19.8 \, \textsf{mag}\). The limiting redshift is helpful for doing volumetric corrections (see section \ref{ssec:methodology_vmax}). 


\subsection{Environmental Densities}
\label{sec:data_env_densities}
We use four different environmental densities in our dataset as provided by GAMA and as described below. The first three estimate the number of galaxies within the local neighbourhood, while the geometric environmental measure is related to the location of the galaxy within the cosmic web. As described in section \ref{sec:intro_env_effects}, the first three are measures of local environment, while the final is a measure of the large-scale environment.

We note that all the local environmental measures are defined on a density defining population (DDP) of galaxies which is also volume limited (see section \ref{ssec:methodology_vmax}). This population
is defined as all galaxies with \(M_r(z_{ref}=0, Q=0.78) < -20 \, \textsf{mag}\), where \(Q\) defines the expected evolution of \(M_r\) as a function of redshift \citep{2012MNRAS.420.1239L}. Given GAMA's limiting magnitude of \(r < 19.8\, \textsf{mag}\), this leads to a redshift limit of \(z \lesssim 0.18\) for these density mesures. See \cite{broughGalaxyMassAssembly2013} for more information.

We note that objects in our sample have
estimates defined for each environmental measure. Also, we exclude any objects that do not have reliable density estimates due to survey edge effects. 

\subsubsection{5th Nearest Neighbour Surface Density (\(\Sigma\))}

This environmental measure estimates the projected 2D surface density based on the distance to the \(N\)-th nearest spectroscopically confirmed bright neighbouring galaxy (within the DDP) within a given redshift and absolute magnitude range. If this distance is \(d_N\), then the surface density is given by 

\begin{equation}
	\Sigma = \frac{N}{\pi d_N^2}
	\label{eq:data_5nn}
\end{equation}

GAMA uses \(N = 5\), a redshift range \(\pm \triangle z c = 1000 \textsf{km/s}\) and absolute magnitude \(M_r < -20\, \textsf{mag}\) \citep{broughGalaxyMassAssembly2013}. When one or more of the \(N\)neighbours are beyond the survey edge, additional considerations are required and upper limits are estimated. However, we do not include any such galaxies in our dataset. 

Figure \ref{fig:data_hist_log_sigma} shows the distribution of \(\Sigma\) for the galaxies in our sample. Note that in this paper we always plot environment measures in logarithmic units, since we are interested in their effects on an order of magnitude scale.

\subsubsection{Cylindrical Count (CC)}
\label{sssec:data_cc}
Similar to the 5th nearest neighbour surface density, this measure provides the number of (other) galaxies (within the DDP) within a cylindrical volume centered on the galaxy in question. It is given by

\begin{equation}
	CC = \frac{N_{cyl}}{\bar{n}_{ref} V_{cyl}} \propto N_{cyl}
\end{equation}
where \(N_{cyl}\) is the number of galaxies in the cylinder and \(V_{cyl}\) is the volume of the cylinder. The (co-moving) radius of the cylinder is taken as 1 Mpc, and the height is determined by the redshift range of \(\pm 1000 \textsf{km/s}\) as before. \(\bar{n}_{ref} = 0.00911 Mpc^{-3}\) is the average number density of the density defining population. See \cite{liskeGalaxyMassAssembly2015} for more information. Since \(V_{cyl}\) and \(\bar{n}_{ref}\) are the same for every object, we can use the galaxy count \(N_{cyl}\) as a proxy for the overdensity.

Figure \ref{fig:data_hist_log_cc} shows a histogram of the cylindrical counts. About \(25 \%\) of the objects do not have any galaxies in their cylinder and so have \(N_{cyl} = 0\) 
In order to include these zero counts when working in log units, we set \(\log(CC) = -1\) when \(CC = 0\), which is why the histogram has a huge bar centered on -1.



\subsubsection{Adaptive Gaussian Environment Density (AGE)}

This density measure is equivalent to a weighted local volume density of galaxies, where closer galaxies receive more weight than more distant ones \citep{yoonSpectrophotometricSearchGalaxy2008}. First, the neighbouring galaxies are identified (from the DDP) as those lying within an adaptive Gaussian ellipsoid defined by

\begin{equation}
	\left(\frac{r_a}{3\sigma}\right)^2 + \left(\frac{r_z}{\textsf{AGEScale} * 3 \sigma} \right)^2 \leq 1
\end{equation}
where \(r_a, r_z\)  are the distances from the centre of the ellipsoid (i.e. from the position of the galaxy in question) in the plane of sky and along the line-of-sight in co-moving Mpc, respectively, and \(\sigma = 2 \textsf{Mpc}\). AGEScale is the adaptive scaling factor, defined as

\begin{equation}
	\textsf{AGEScale} = 1 + (0.2\,n)
\end{equation}

where \(n\) is the number of galaxies from the DDP within 2 Mpc. \textsf{AGEScale} is used to scale the value of sigma along the redshift axis by up to a factor of 3 (corresponding to \(n = 10\)) for the highest density environments to compensate for the "finger-of-God" effect \citep{schawinskiEffectEnvironmentUltraviolet2007,thomasEnvironmentSelfregulationGalaxy2010}. The adaptive Gaussian density parameter is then computed as:

\begin{equation}
	AGE = \frac{1}{\sqrt{2  \pi} \sigma} \sum_i \exp \left[ - \frac{1}{2} \left( \left(\frac{r_{a,i}}{\sigma}\right)^2 + \left(\frac{r_{z,i}}{\textsf{AGEScale} * \sigma} \right)^2 \right) \right]
\end{equation}

Figure \ref{fig:data_hist_log_age} shows a histogram of the Adaptive Gaussian Environment density.

 

\subsubsection{Geometric Environment: GeoS10}

This environmental measure identifies the cosmic web of large scale structure within the GAMA equatorial survey regions by classifying the geometric environment of each point in space as either a void, a sheet, a filament or a knot \citep{eardleyGalaxyMassAssembly2015}.
The label is assigned based on the number of dimensions along which the underlying matter is collapsing. Galaxies are used as a proxy to estimate the overall matter distribution. A Hessian (matrix of second-order partial derivatives) of the gravitational potential is computed, and eigenvalues of the Hessian give the directions along which the Hessian is changing. This matrix is also called the tidal tensor. The number of positive eigenvalues is the number of dimensions along which matter is collapsing. In practice, the only directions considered are ones where the collapse is happening reasonably quickly. This is equivalent to counting the number of eigenvalues higher than a certain threshold. The classification assigned is as follows :

\begin{itemize}[labelindent=0pt,labelwidth=0pt, labelsep*=0pt, leftmargin=!, style=standard]
	\item[] \textbf{Voids}:  all eigenvalues below the threshold (\(\lambda_1 < \lambda_{th}\))
	\item[] \textbf{Sheets}: one eigenvalue above the threshold (\(\lambda_1 > \lambda_{th}, \lambda_2 < \lambda_{th} \)).
	\item[] \textbf{Filaments}: two eigenvalues above threshold (\(\lambda_2 > \lambda_{th}, \lambda_3 < \lambda_{th}\)).
	\item[] \textbf{Knots}: all eigenvalues above the threshold (\(\lambda_3 > \lambda_{th}\)).
\end{itemize}

where \(\lambda_1, \lambda_2, \lambda_3\) are the three eigenvalues, with \(\lambda_3 < \lambda_2 < \lambda_1\). \(\lambda_{th}\) is the threshold mentioned above. Thus, a `void' region means that there is no significant collapse happening along any direction -- it is a region surrounded by similar or higher density regions in all directions. Similarly, a `sheet' region means that matter is collapsing mainly along a single direction, and so on.

The original galaxy counts used to estimate the matter distribution are smoothed to minimize noise and remove non-linearities. The smoothing scale \(\sigma\) is also a free parameter in addition to the eigenvalue threshold \(\lambda_{th}\). \cite{eardleyGalaxyMassAssembly2015} chose values of \(\sigma\) and \(\lambda_{th}\) so as to divide the galaxies into the four regions (determined by the four labels above) as equally as possible, so that any measurements made from these maybe statistically significant. Two combinations of these parameters are chosen: \((\sigma, \lambda_{th}) = (4 h^{-1} \textsf{Mpc}, 0.4)\) or \((10 h^{-1} \textsf{Mpc}, 0.1)\). Both combinations have at least 10\% of galaxies within each of the four regions. For our work, we want to use this measure to study environmental effects on large scales and be independent of the local environmental measures. Hence we choose the estimates from the larger smoothing scale of \(10\, h^{-1}\, \textsf{Mpc}\). We refer to it as GeoS10 in accordance with \cite{eardleyGalaxyMassAssembly2015}. 
Figure \ref{fig:data_hist_geos10} shows the histogram of this environmental measure. Overall there is a statistically significant number of galaxies in each of the four bins, with filaments have the highest (\(\sim 30,000\)) and voids having the lowest (\(\sim 7500\)) number of galaxies. 

The definition of geometric galaxy environment employed in this work partitions space into regions defined as being voids, sheets, filaments or knots. Galaxies are labeled according to the region they reside in, e.g. a galaxy within a knot region is defined as a knot galaxy (see \citealt{eardleyGalaxyMassAssembly2015}, their Figs.~2 and 3). The geometric environment is therefore discretely defined. Other methodologies, such as the \textsc{DisPerSE} algorithm \citep{sousbiePersistentCosmicWeb2011}, directly construct cosmic webs such that they are skeleton structures that exist within space, rather than partitioning it directly. Continuous studies of environment, such as how galaxy properties vary with distance to the closest knot \citep[e.g.][]{kraljicGalaxyEvolutionMetric2018,malavasiRelativeEffectNodes2022}, are then possible. An extension of the \citet{eardleyGalaxyMassAssembly2015} methodology to enable such studies here are beyond the scope of this work.

\section{Methodology}
\label{methodology}

\begin{figure*}
	\centering
	\begin{adjustbox}{width=1.1\textwidth,center=\textwidth}	
		\begin{subfigure}[b]{\columnwidth}   
			\centering 
			\includegraphics[width=\textwidth]{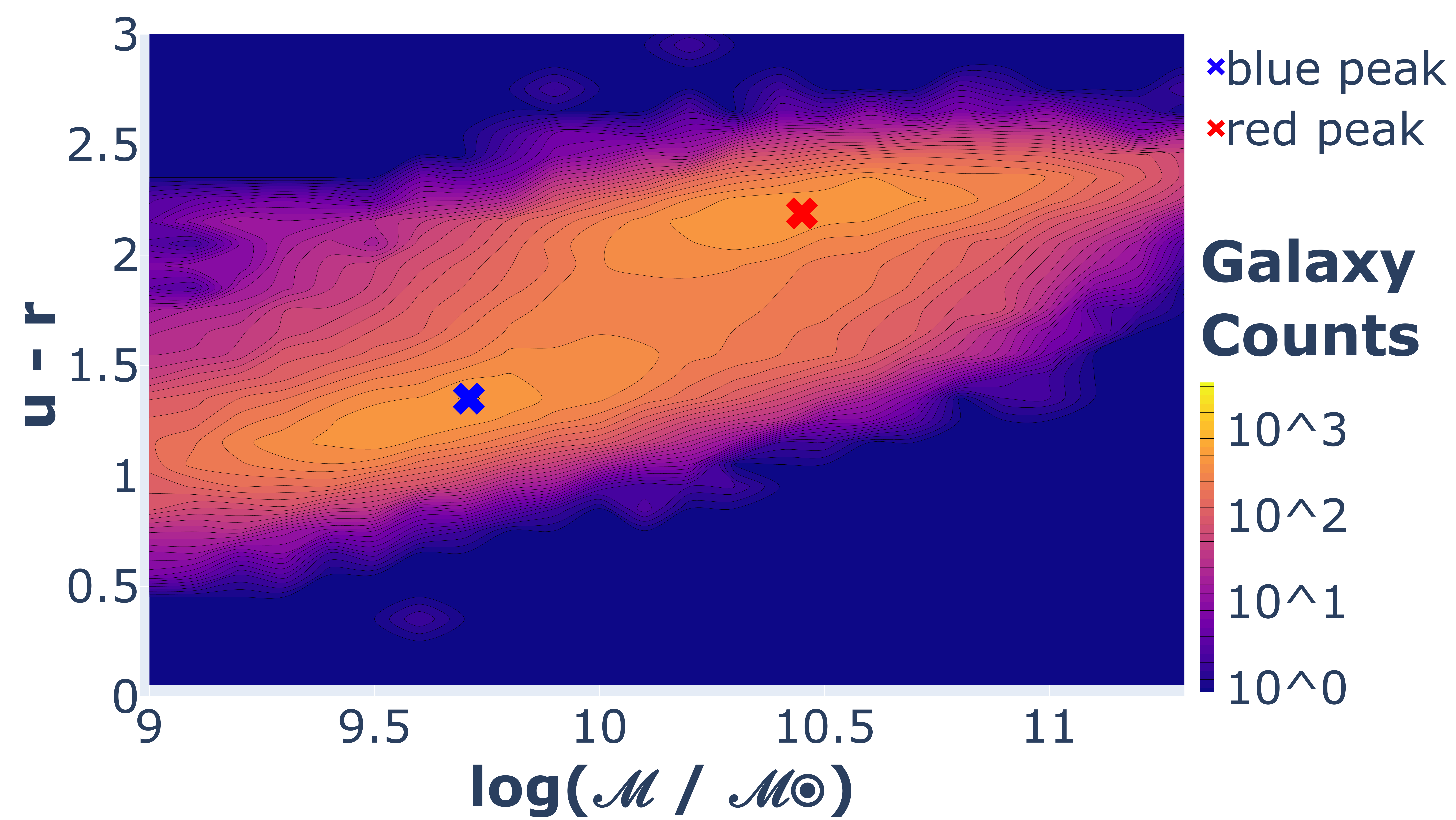}
			\caption[]%
			{{\small Without volumetric weighting}}    
			\label{fig:methodology_contour_uminusr_non_vmax}
		\end{subfigure}
		\hfill
		\begin{subfigure}[b]{\columnwidth}   
			\centering 
			\includegraphics[width=\textwidth]{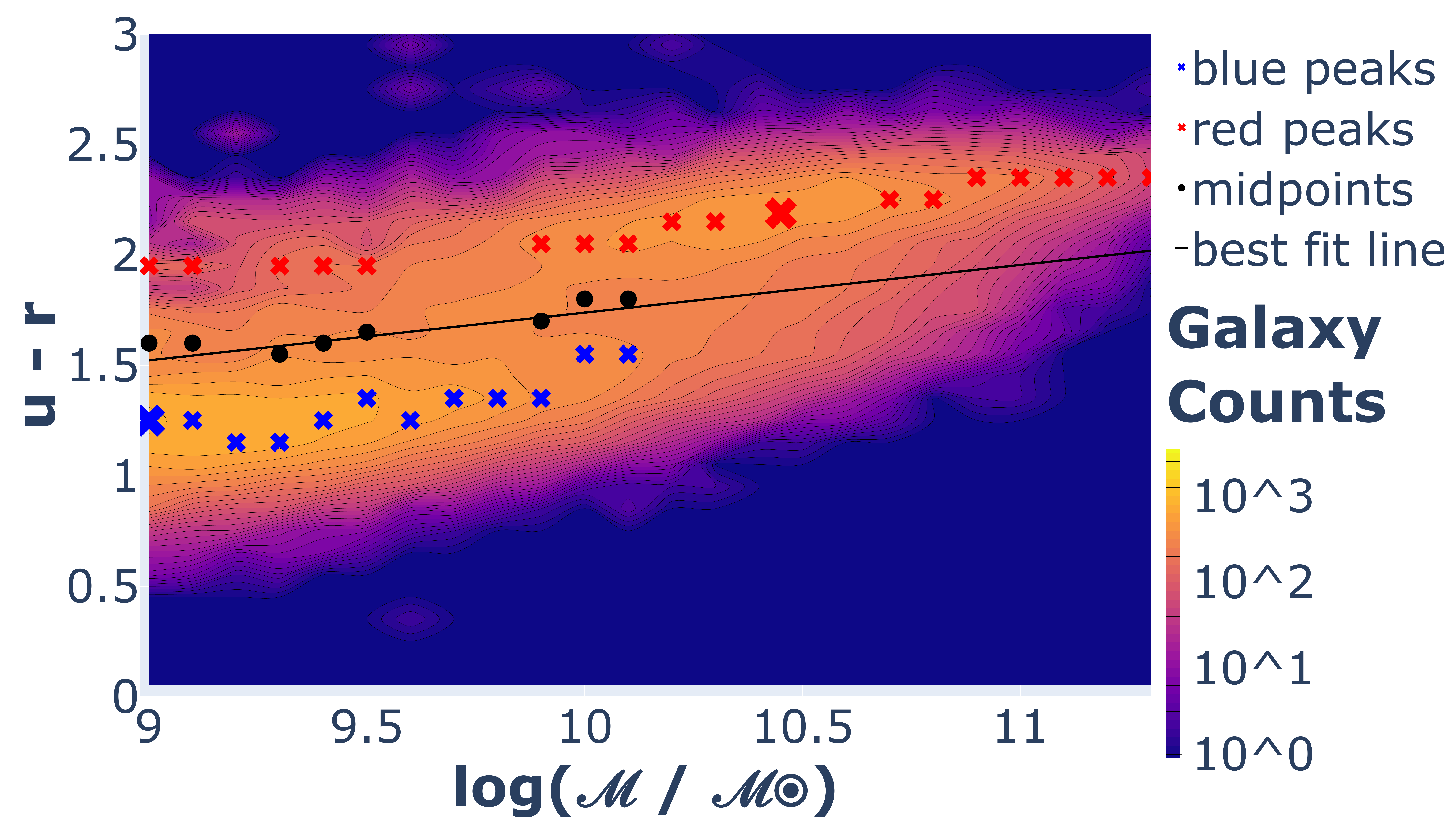}
			\caption[]%
			{{\small volume weighted with dividing line}}    
			\label{fig:methodology_contour_uminusr_peaks_midpoints}
		\end{subfigure}
	\end{adjustbox}
	\captionsetup{width=\textwidth}
	\caption[ ]
	{{ \small Contour plots and heatmaps representing the distribution of galaxy counts as function of the stellar mass \(\log(\mathcal{M}/\mathcal{M_\odot})\) and colour \(u - r\) . Panel (a) shows the raw counts, we see two peaks corresponding to red and blue region, shown as cross marks of the corresponding colour. Panel (b) shows volume (\(1/V_{max}\)) weighted counts. The corresponding peaks from Panel (a) are shown by large cross symbols. It also shows peaks in the red and blue region for each mass bin (represented as dots of the corresponding colour) along with midpoints (black dots) and the best line line of separation (black line). }}
	\label{fig:methodology_contour_plots}
\end{figure*}

Figure \ref{fig:methodology_contour_uminusr_non_vmax} shows a colour mass diagram (contour plot + heatmap) of galaxy counts. 
We can identify two peaks (shown by blue and red crosses in the figure), which would typically be associated with the separation of the galaxy population into a blue and red sequence \citep{tullyColormagnitudeRelationSpiral1982,stratevaColorSeparationGalaxy2001,baldryQuantifyingBimodalColormagnitude2004,baldryColorBimodalityImplications2004,taylorGalaxyMassAssembly2015}. To understand the effect of mass and environment on galaxy colour, we first need to separate the red and blue galaxy populations (section \ref{ssec:methodology_separating_populations}), so that we can estimate the red galaxy fraction and study its evolution. Before we do that, we need make sure that the sample is volumetrically unbiased (section \ref{ssec:methodology_vmax}). 

\subsection{Volumetric Correction}
\label{ssec:methodology_vmax}
\subsubsection{The \(1/V_{max}\) approach}
Volumetric correction is important for our analysis because blue galaxies are brighter at a given stellar mass, and would therefore be overrepresented as compared to the red galaxies. This would lead to biased red fraction estimates.

 We follow the approach of \cite{feltenSchmidtEstimatorOther1976} to perform volumetric correction. Intuitively, as we observe out to higher redshifts, we will necessarily underestimate the number of faint galaxies, since the light reaching us from them is not sufficient for inclusion in the survey. The magnitude of a galaxy determines the redshift range over which it can be observed, and there may be a number of galaxies whose observable redshift range is much smaller than the survey redshift range. To correct for this, we weigh each galaxy by a factor \(f = \frac{V_{survey}}{V_{max}}\) to make the sample complete (unbiased) within the volume, i.e. the chosen sky area and redshift limits. The term \(V_{max}\) is the maximum comoving volume over which the galaxy under concern can be observed within the redshift limits of the sample. For a galaxy, \(i\), we compute \(V_{max, i}\) as

\begin{equation}
	V_{max, i} \propto [ D(z_{max, i})^3 - D(z_{min,i})^3 ]
\end{equation}
 
 \(D(z)\) is the comoving distance to redshift z 
 . \(z_{max, i}\) is the limiting redshift of the galaxy \(i\) (see section \ref{ssec:data_redshift}) at or the upper limit of the sample, whichever is lower. \(z_{min, i} = z_{start}\) is the lower redshift limit of the sample.

The survey volume is computed in a similar manner
\begin{equation}
	V_{survey} \propto [ D(z_{limit})^3 - D(z_{start})^3 ]
\end{equation}
where \(z_{limit}\) and \(z_{start}\) are the upper and lower redshift limits of the sample. 

To estimate the limiting redshift (\(z_{max}\)), we use the distance modulus equation. Ignoring the evolution of the galaxy over the redshift range, we can write

\begin{align}\nonumber
	m - \mu(z) - K(z) &= m_{limit} - \mu(z_{max}) - K(z_{max})\\
	\mu(z_{max}) &= \mu(z) + [m_{limit} - m] + [K(z) - K(z_{max})]
\end{align}
which can be used to determine \(z_{max}\). \(\mu = 5 \log_{10}(D_L/10 pc)\) is the distance modulus (\(D_L\) is the luminosity distance), \(K(z)\) is the K-correction. The differential K-correction \(K(z_{max}) - K(z)\) is usually small and can be determined from the galaxy SED.

For our chosen sample, we have \(z_{start} = 0.05, z_{limit} = 0.18\). We use the values of maximum redshifts \(z_{max}\) as provided by GAMA for a limiting Petrosian \(r\)-band magnitude of \(19.8 \, \textsf{mag}\), i.e. \(m_{limit} = 19.8\, \textsf{mag}\). These values are available under the column name \(\texttt{zmax\_19p8}\) in the \(\texttt{STELLARMASSES}\) DMU \citep{driverGalaxyMassAssembly2022}.

The \(1/V_{max}\) is a straightforward approach for working with data that is not volumetrically complete. It does however implicitly assume that the distribution of galaxies varies uniformly with redshift. Modifications to the method have been proposed to deal with the presence of large-scale structure, e.g. by introducing additional terms in the weighting factor \(V_{survey}/V_{max}\) (see \citealt{baldryGalaxyMassAssembly2012}).

\subsubsection{Clipping outliers}

Photometric errors can sometimes result in incorrect estimates of the limiting redshift, causing outliers to be present in the \(z_{max}\) distribution at a fixed mass. The outliers at lower values of \(z_{max}\) can cause more problems, since they would result in a higher values of \(1/V_{max}\), thereby skewing up the overall galaxy counts. To avoid these outliers, we clip the \(z_{max}\) distribution to the 95th percentile within each (log) mass bin of width 0.01. The percentiles are computed for bins spaced at intervals of \(0.25\) and a second order polynomial curve is fitted to these points:

\begin{equation}
	f(x)= a_0 + a_1 x + a_2 x^2
\end{equation}

The best fit values obtained are \((a_0, a_1, a_2) = (5.03, -1.1, 0.06)\). 


The \(1/V_{max}\) weights have values ranging from 1 -- which is the median and the most common weight, to 18.6 -- which occurs for low-mass galaxies (\(\mathcal{M} \approx 10^9 \mathcal{M_\odot}\)) The mean weight and standard deviation are both about 1.6. 

Figure \ref{fig:methodology_contour_uminusr_peaks_midpoints} shows the updated galaxy counts as a function of colour and stellar mass, corrected for volumetric completeness. We can see that while the red peak has stage at its original place, the blue peak has move towards lower mass \(\mathcal{M} \approx 9 \mathcal{M}_\odot\). Also, the correlation between mass and colour has reduced for the region around the blue peak, as compared to Figure \ref{fig:methodology_contour_uminusr_non_vmax}.

\subsection{Separating the two populations}
\label{ssec:methodology_separating_populations}

We can now use the \(1/V_{max}\) weighted galaxy counts to separate the red and blue populations. We assume that the colour evolution of the population over our redshift range of interest is not significant enough to matter for the separation. The simplest approach for this would be to fit a line through the contour distribution (equivalently a plane through the 3D distribution) that divides the two populations well. For our purposes, it is not necessary for this line to be exact - an approximate line would add some noise when computing the red fraction within a given mass/environment bin, but it would still retain the overall signal of the variation in red fraction varies as a function of stellar mass and environmental density. 

 We first bin the weighted galaxy counts by mass, each bin being of width \(\triangle \log (\mathcal{M} / \mathcal{M_\odot}) = 0.1\). Within each mass bin, we assume the counts are bimodally distributed as a function of colour, similar to \cite{baldryQuantifyingBimodalColormagnitude2004}. We therefore search for two peaks (local maxima) within each bin - one each for the red and blue populations. We restrict our search for the blue peak to \(0 < u - r < 1.5\) and for the red peak to \(1.5 < u - r < 2.25\) - the upper limit of 2.25 is to avoid spurious local maxima that may falsely get identified as real peaks.
We then use the midpoint of the two peaks as the separating colour value for that bin. Fitting a line through these midpoints gives us our line of separation for the two populations. The equation of the line of separation used is

\begin{equation}
	C = 0.216  M - 0.421
\end{equation}

where \(C\) is the colour \(u - r\) and \(M\) is the galaxy stellar mass in \(\log (\mathcal{M}/\mathcal{M_\odot})\). Figure \ref{fig:methodology_contour_uminusr_peaks_midpoints} shows the results of this approach. 
The red and blue peaks are identified as dots of the corresponding colour. The red peaks seem to fall on a straight line with limited scatter, and align well with the contour lines. The blue peaks do the same for lower masses \(\mathcal{M} \lesssim 10^{10} \mathcal{M_\odot}\). The contours seemed to flatten out in the range \(\mathcal{M} \gtrsim 10^{10} \mathcal{M_\odot}, 0 < u - r < 1.5\) and there are hardly any peaks here. 
We found that varying the different peaks changed the midpoints and the separating line but did not affect the environment signal mentioned above, so we maintain this simple approach. The midpoints and the separating line are shown in black in Figure \ref{fig:methodology_contour_uminusr_peaks_midpoints}.

\section{Results and Discussion}
\label{results_discussion}

\begin{figure*}
	\centering
		\begin{subfigure}[b]{0.498\textwidth}   
			\centering 
			\includegraphics[width=\textwidth]{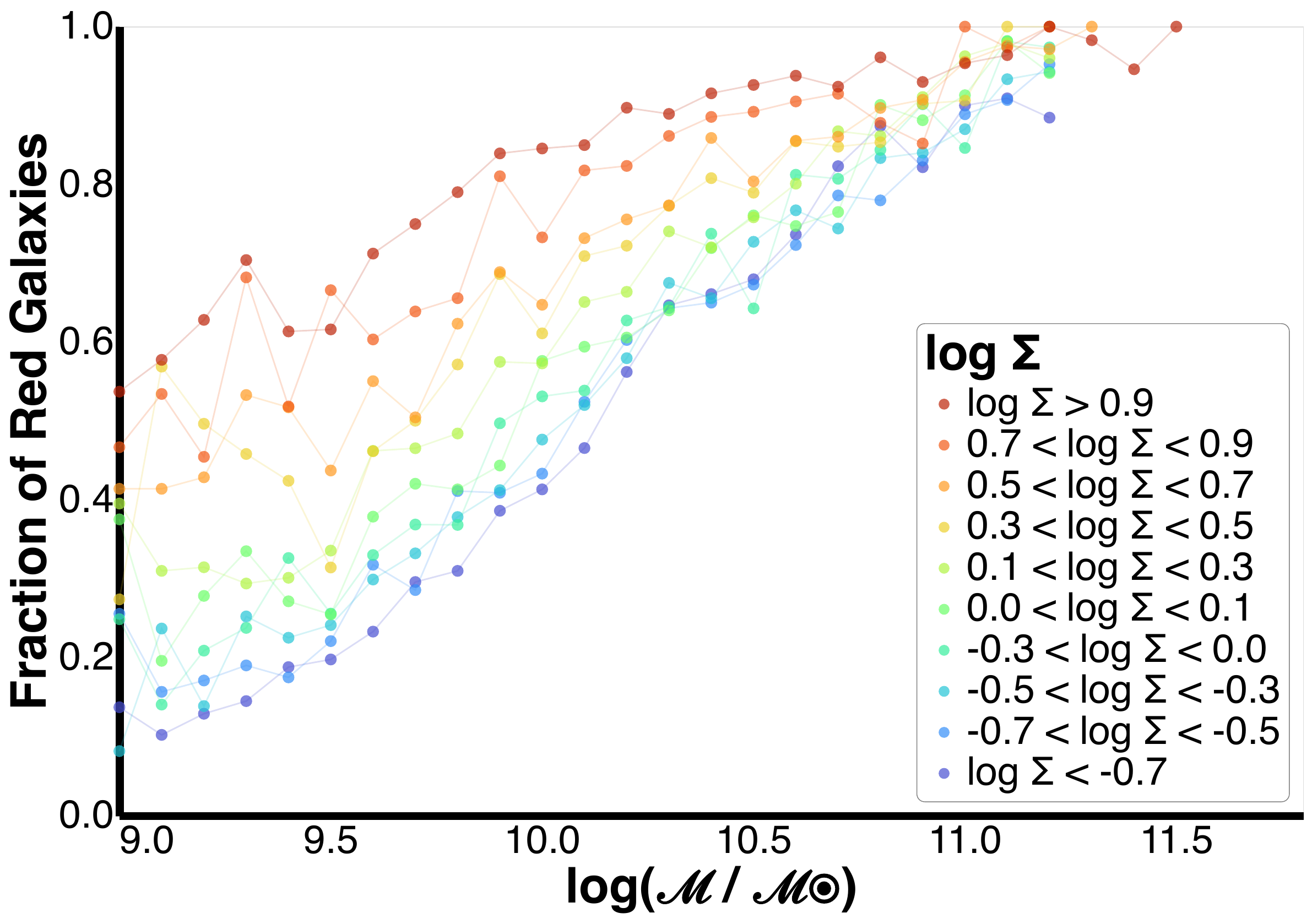}
			\caption[]%
			{{\small}} 
			\label{fig:discussion_chart_5nn}
		\end{subfigure}
		\hfill
		\begin{subfigure}[b]{0.498\textwidth}   
			\centering 
			\includegraphics[width=\textwidth]{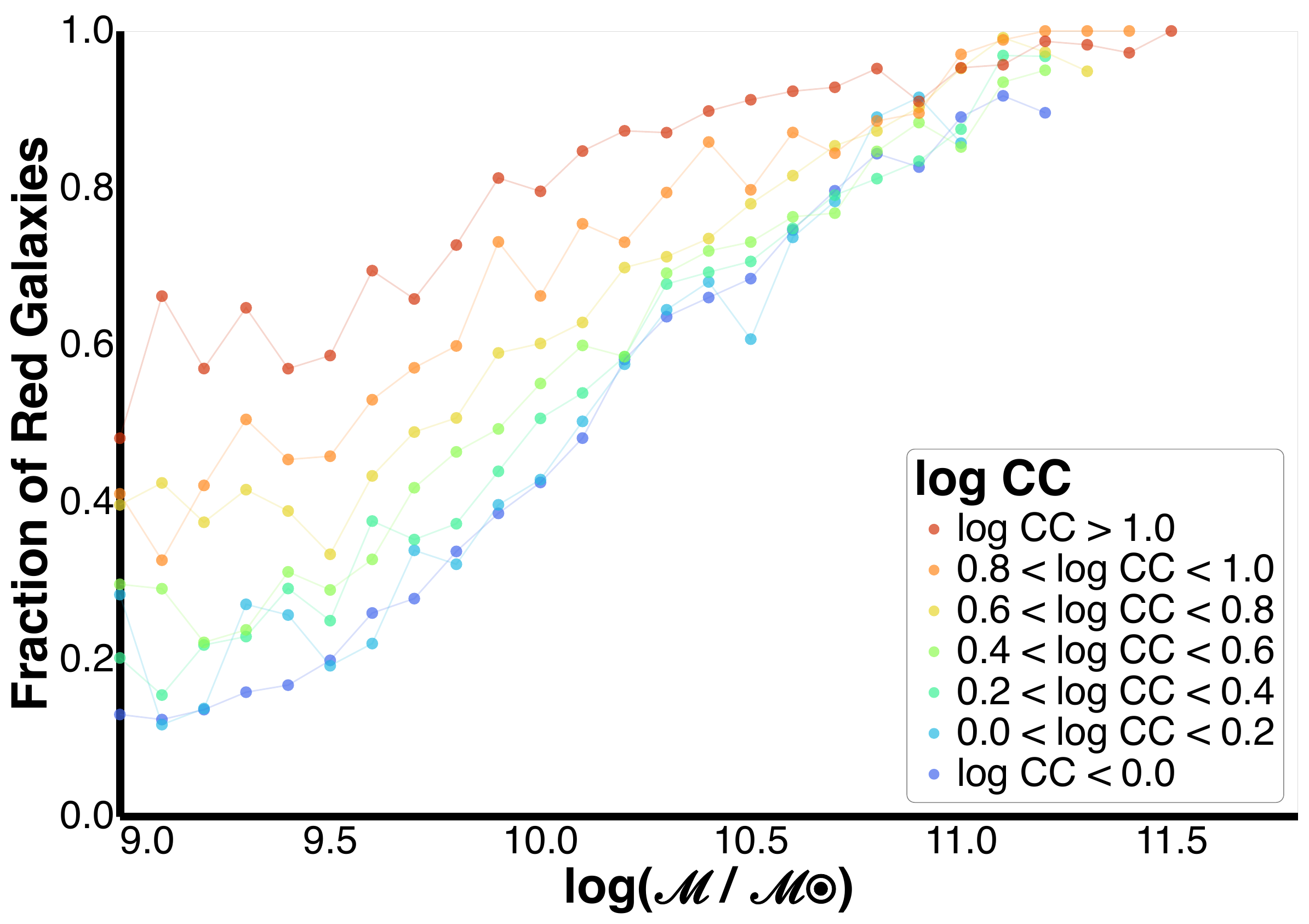}
			\caption[]%
			{{\small}} 
			\label{fig:discussion_chart_cc}
		\end{subfigure}
	\vskip\baselineskip
		\begin{subfigure}[b]{0.498\textwidth}   
			\centering 
			\includegraphics[width=\textwidth]{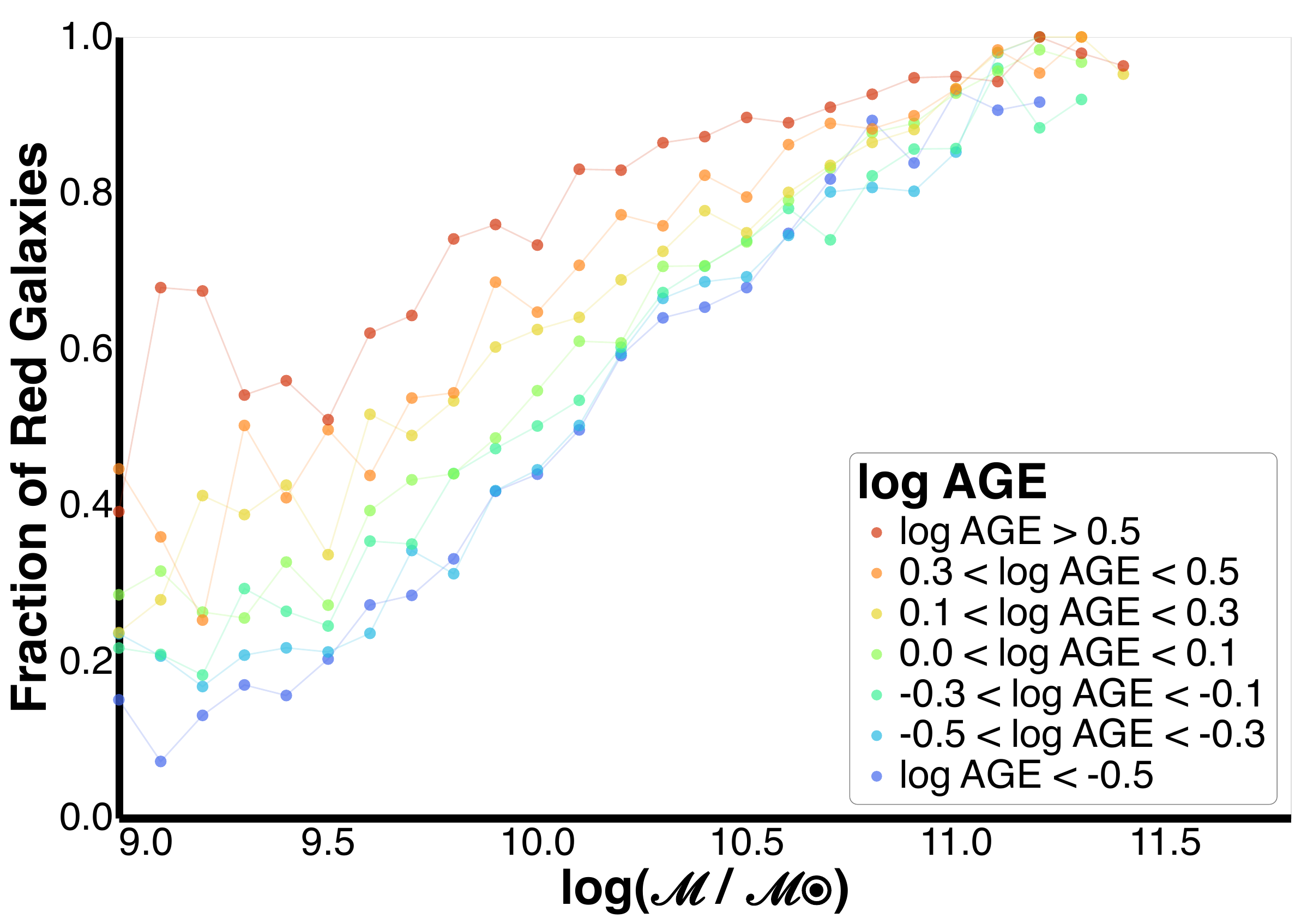}
			\caption[]%
			{{\small}} 
			\label{fig:discussion_chart_age}
		\end{subfigure}
		\hfill
		\begin{subfigure}[b]{0.498\textwidth}   
			\centering 
			\includegraphics[width=\textwidth]{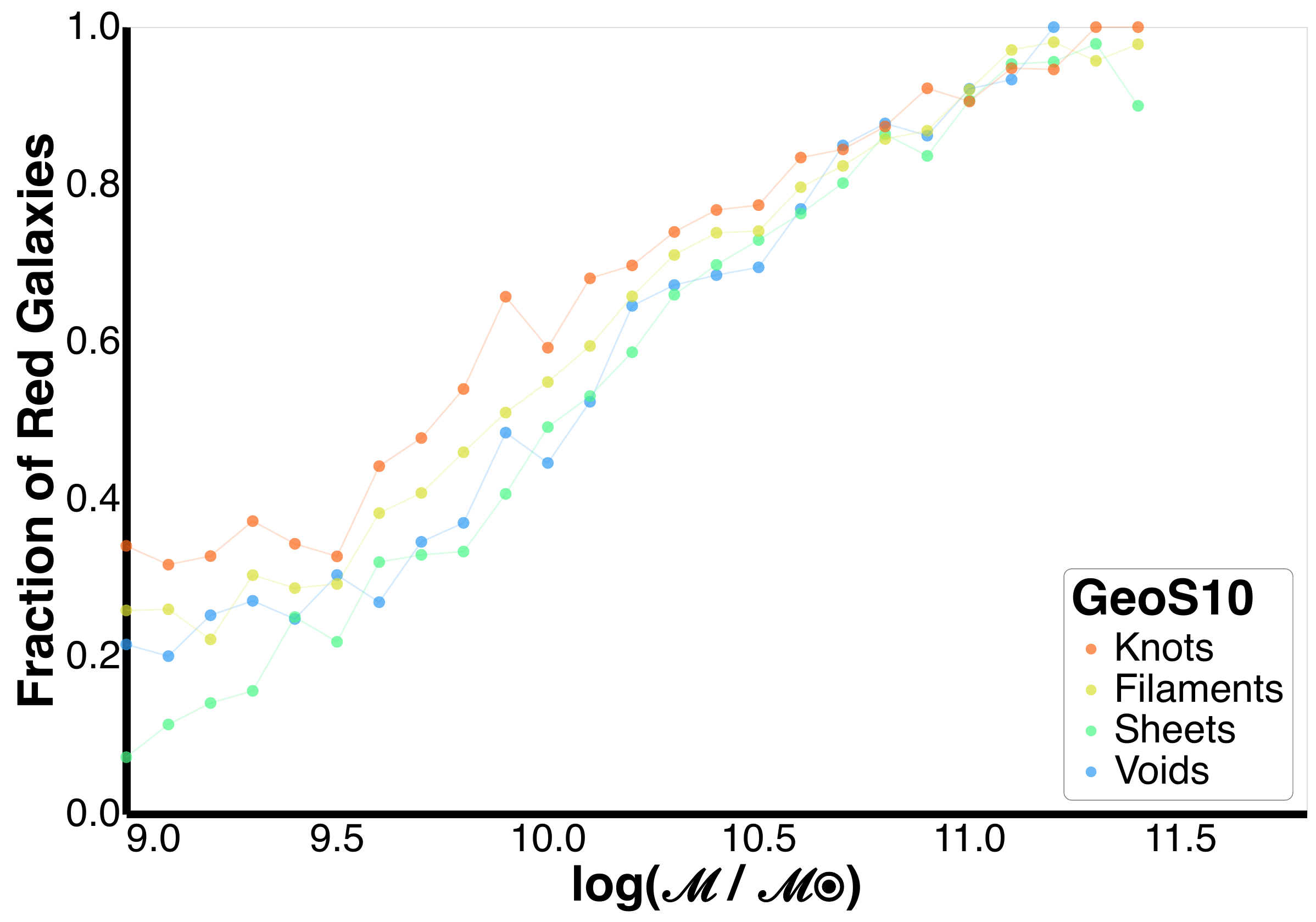}
			\caption[]%
			{{\small}} 
			\label{fig:discussion_chart_geos10}
		\end{subfigure}
	\caption[ ]
	{{ \small Fraction of red galaxies as a function of stellar mass and different environmental densities. The fractions are computed for bins of stellar mass (\(\log (\mathcal{M}/\mathcal{M_\odot})\), shown on the x-axis) and environment (shown by different colours). The plots are for (a) 5th Nearest Neighbour (\(\Sigma\)), (b) Cylindrical Count (CC), (c) Adaptive Gaussian Ellipsoid (AGE) and (d) Geometric Environment (GeoS10). Note the logarithmic scale for both mass and environmental density.
			}}
	\label{fig:discussion_1d_plots}
\end{figure*}

\begin{figure*}
	\centering
	\begin{adjustbox}{width=\textwidth,center=\textwidth}	
		\begin{subfigure}[b]{0.475\textwidth}   
			\centering 
			\includegraphics[width=\textwidth]{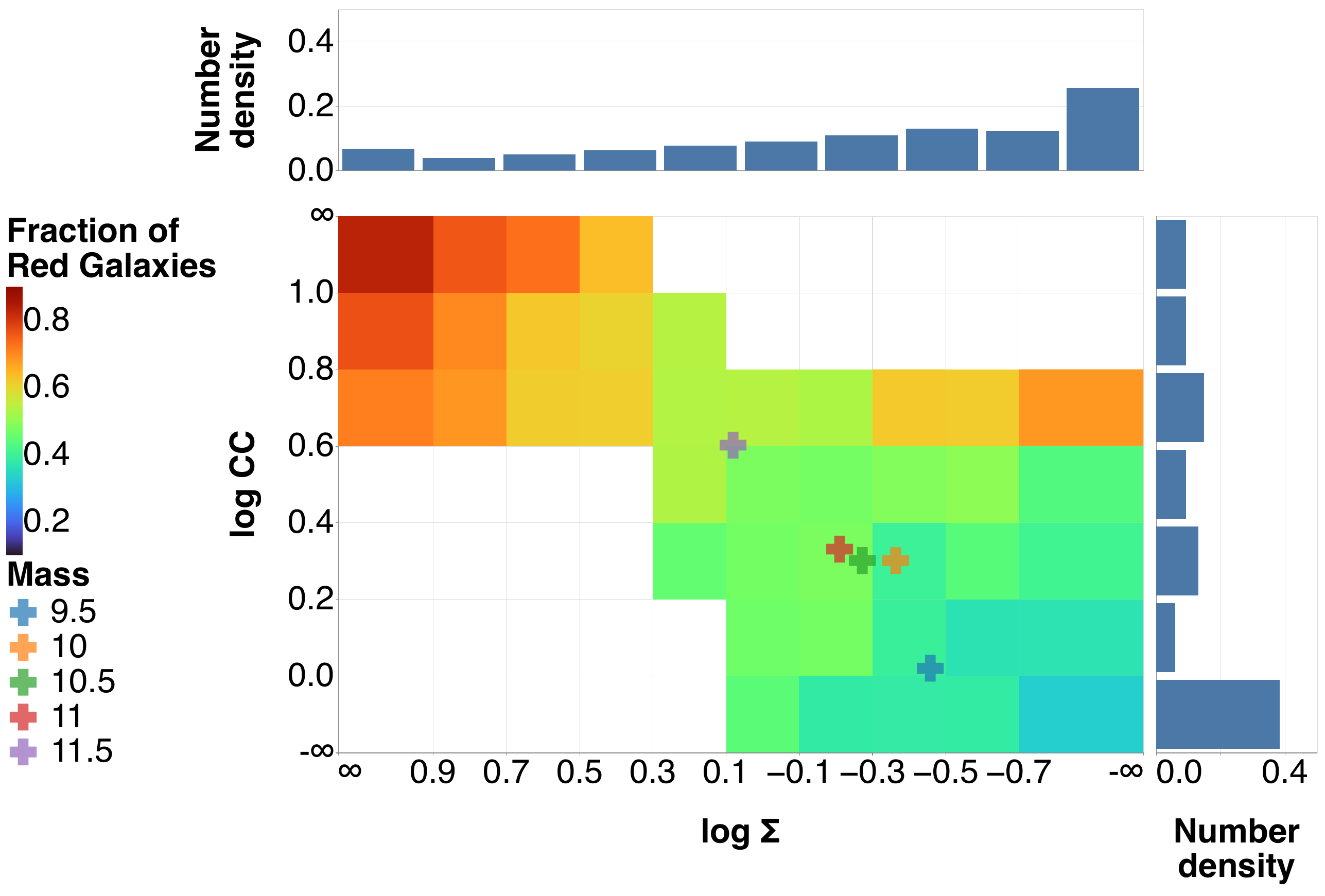}
			\caption[]%
			{{\small \(\Sigma\) vs CC }}    
			\label{fig:discussion_chart_5nn_cc}
		\end{subfigure}
		\hfill
		\begin{subfigure}[b]{0.475\textwidth}   
			\centering 
			\includegraphics[width=\textwidth]{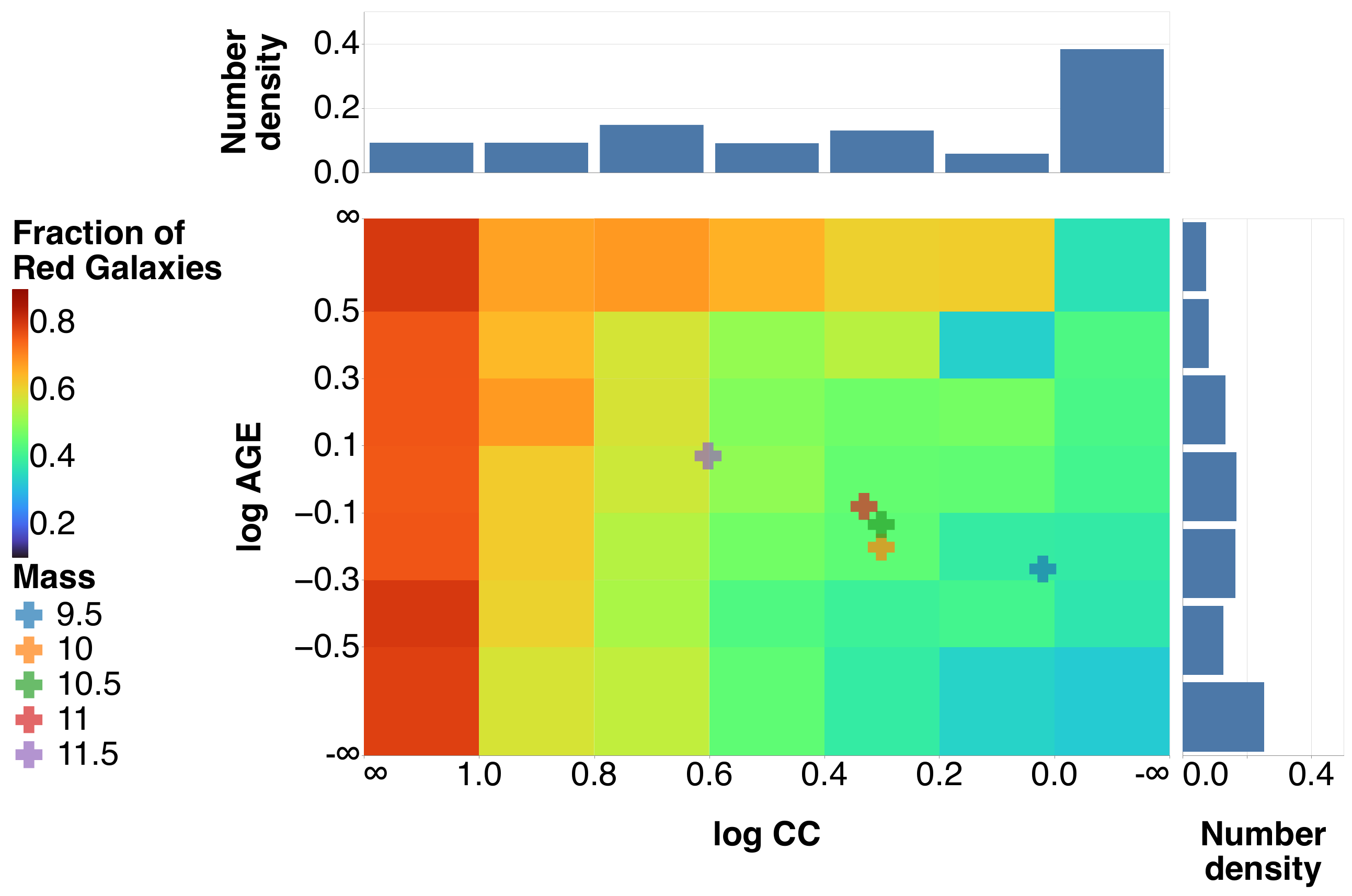}
			\caption[]%
			{{\small CC vs AGE}}    
			\label{fig:discussion_chart_cc_age}
		\end{subfigure}
	\end{adjustbox}
		\vskip\baselineskip
	\begin{adjustbox}{width=\textwidth,center=\textwidth}	
		\begin{subfigure}[b]{0.475\textwidth}   
			\centering 
			\includegraphics[width=\textwidth]{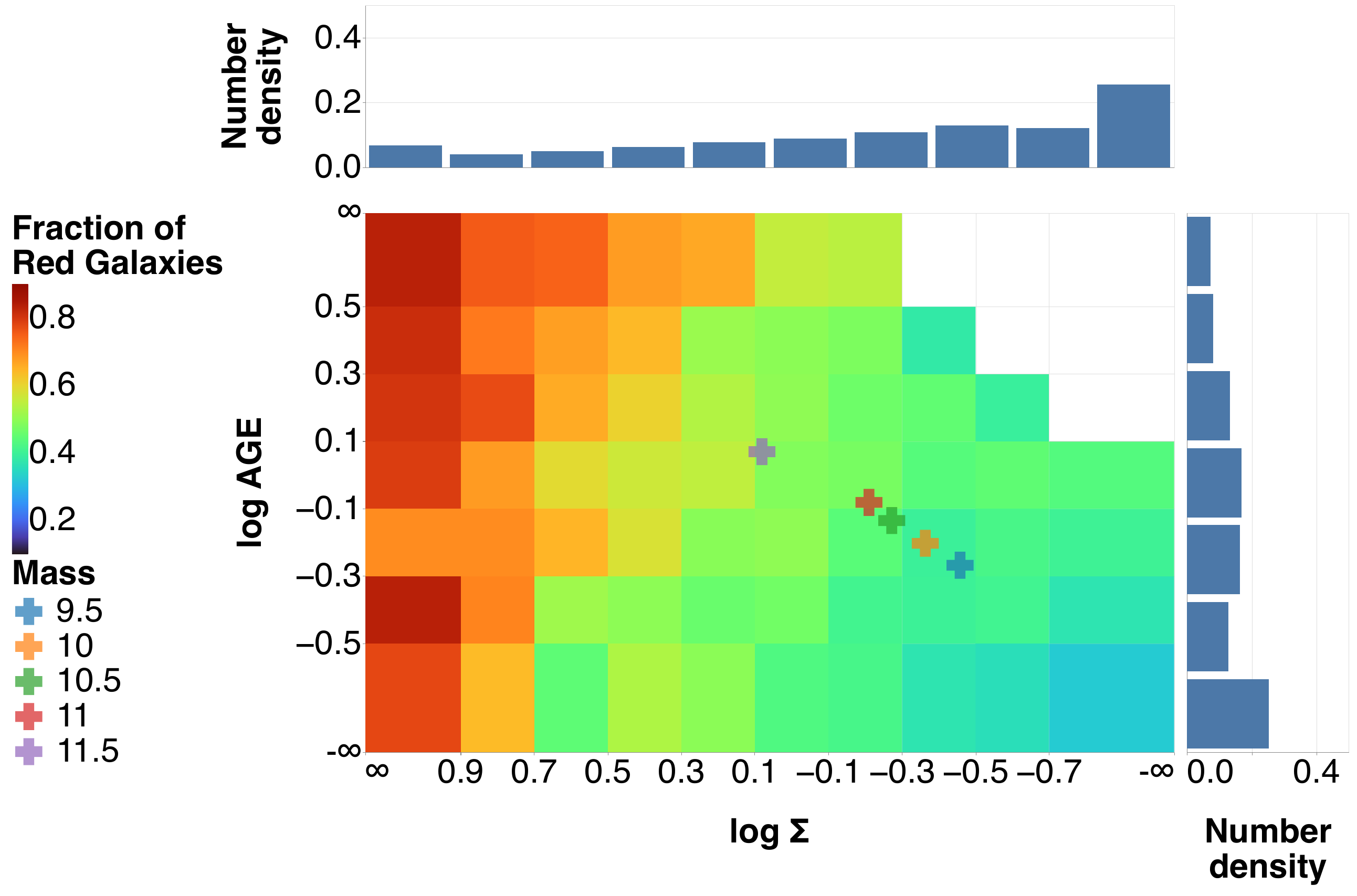}
			\caption[]%
			{{\small \(\Sigma\) vs AGE}}    
			\label{fig:discussion_chart_5nn_age}

		\end{subfigure}
 
	\hfill
		\begin{subfigure}[b]{0.475\textwidth}   
			\centering 
			\includegraphics[width=\textwidth]{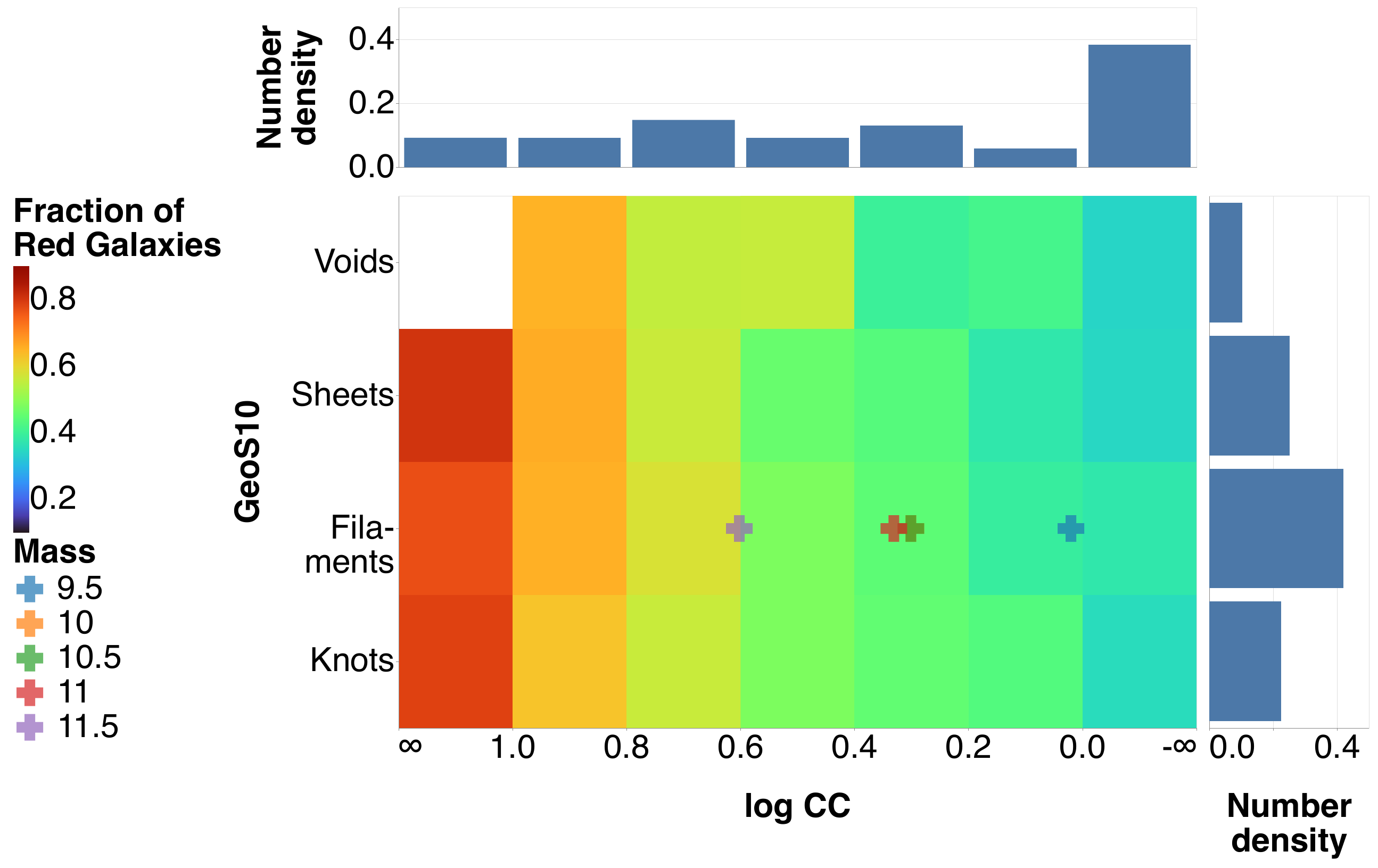}
			\caption[]%
			{{\small CC vs GeoS10}}    
			\label{fig:discussion_chart_cc_geos10}
		\end{subfigure}
		\end{adjustbox}
		\vskip\baselineskip
	\begin{adjustbox}{width=\textwidth,center=\textwidth}	
		\begin{subfigure}[b]{0.475\textwidth}   
			\centering 
			\includegraphics[width=\textwidth]{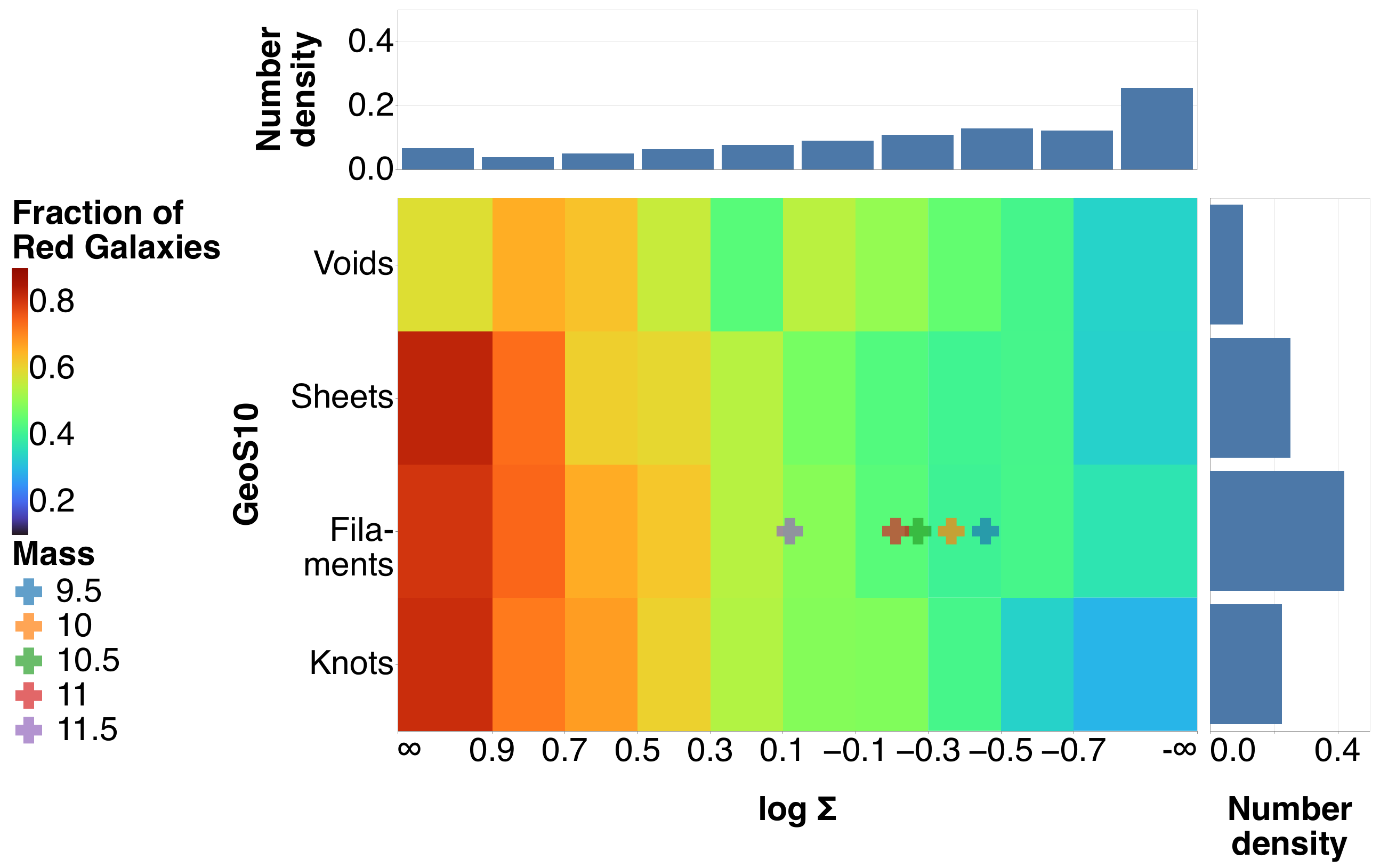}
			\caption[]%
			{{\small \(\Sigma\) vs GeoS10 }} 
			\label{fig:discussion_chart_5nn_geos10}
		\end{subfigure}
		\hfill
		\begin{subfigure}[b]{0.475\textwidth}   
			\centering 
			\includegraphics[width=\textwidth]{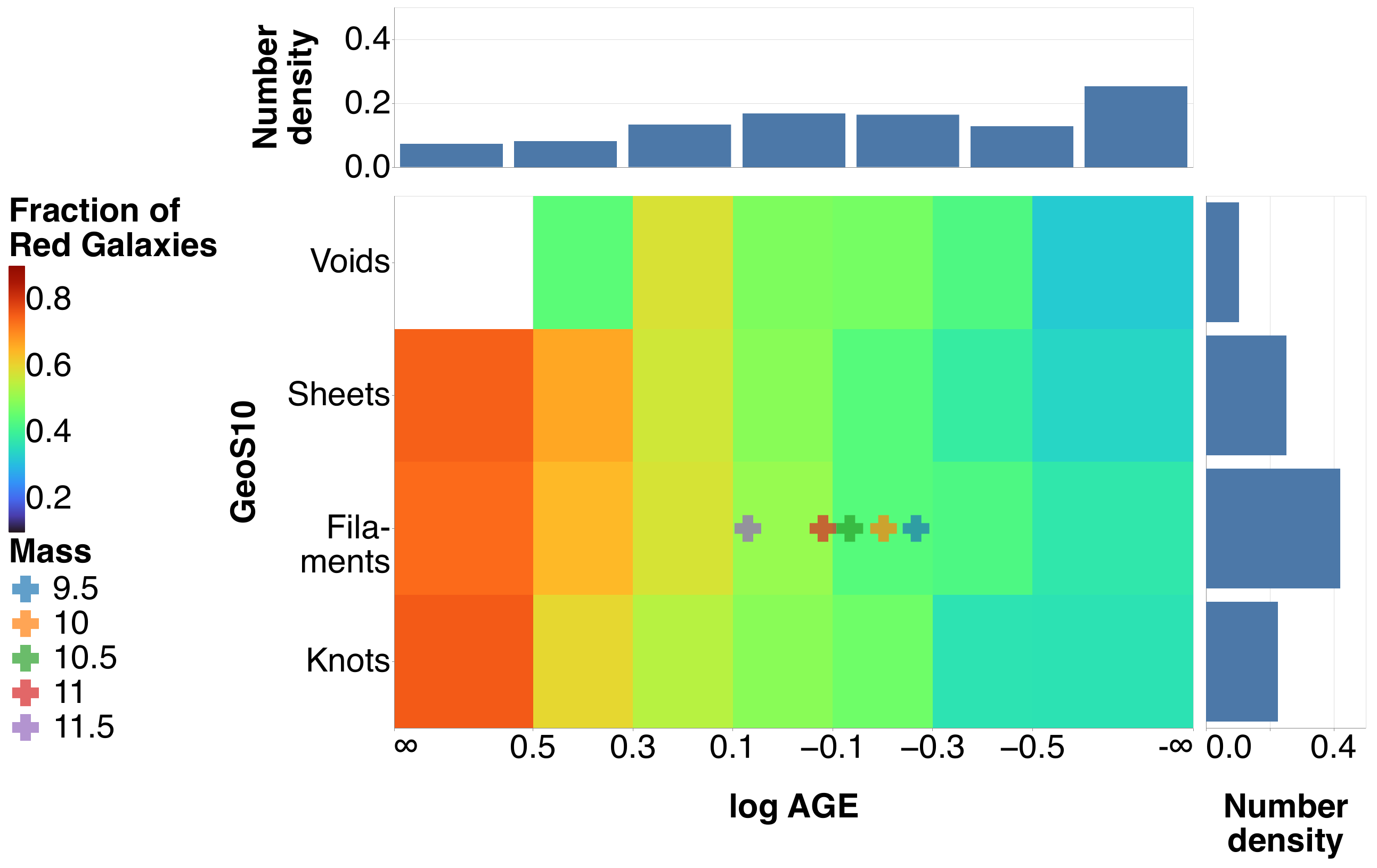}
			\caption[]%
			{{\small AGE vs GeoS10}}    
			\label{fig:discussion_chart_age_geos10}
		\end{subfigure}
	\end{adjustbox}
	\caption[ ]
	{{ \small Heatmaps showing the variation in the fraction of red galaxies (represented as colour) as a function of two different environmental densities (one on each axis). Both environmental densities are binned on a logarithmic scale. White cells indicate regions with \(< 20 \) samples and where a statistically significant estimate was not achievable. Overlaid on the heatmaps are the median (mode for geometric environment) environmental density values for galaxies in different mass bins. These points show that there is an effect on mass that needs to be controlled for, and we do that in the subsequent analysis (section \ref{ssec:discussion_test_effect_geom_env} ). Finally, the sides of each plot show marginalized histograms of individual environmental density, to give information of number density of the galaxies in the different bins. }}
	\label{fig:discussion_2d_plots}
\end{figure*}

\begin{figure}
	\centering
	\includegraphics[width=0.5\textwidth]{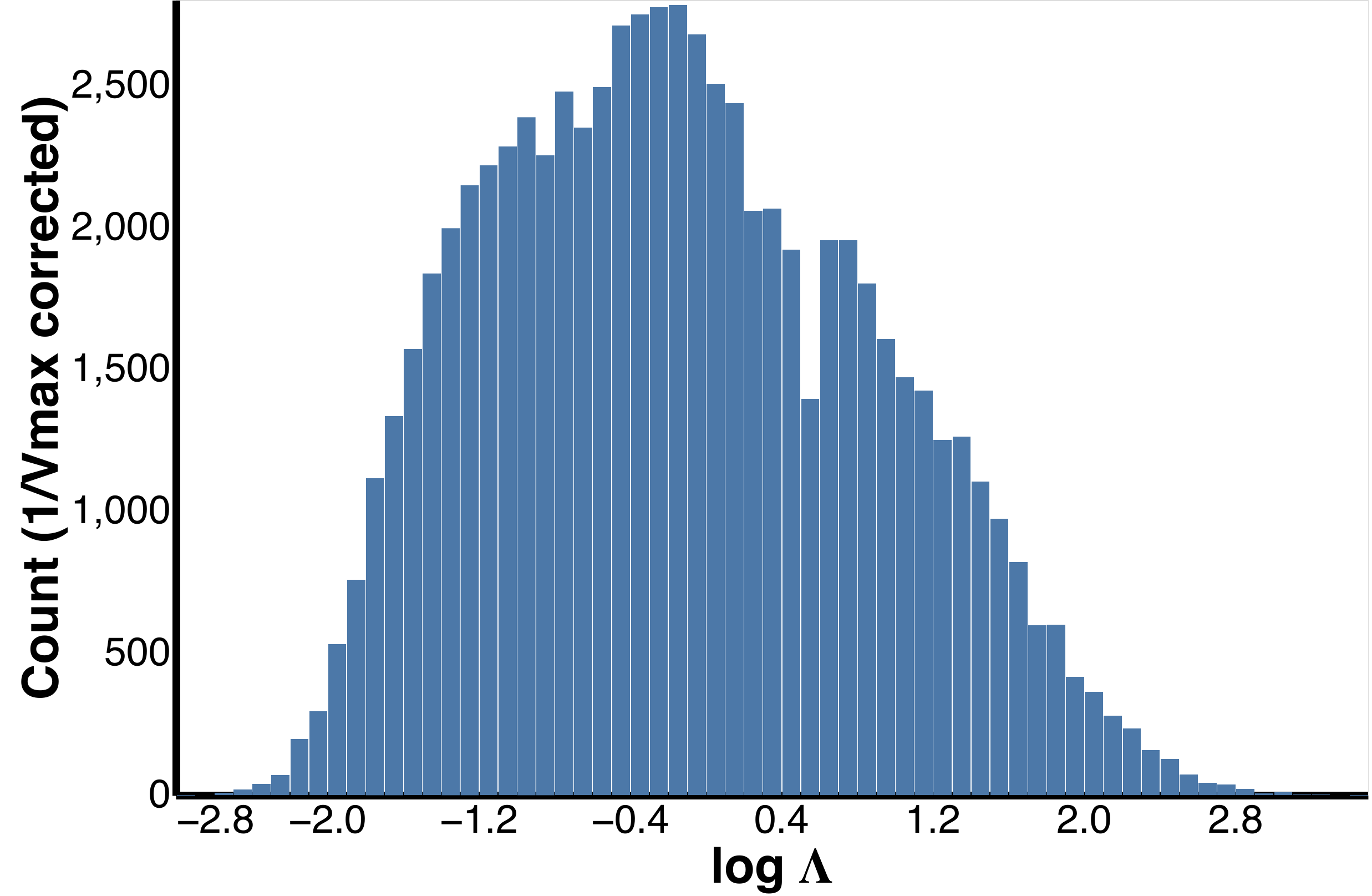}
	\caption{Histogram of local optimal density \(\Lambda\). We model it as a linear combination of local densities (in log space) that maximizes the red fraction range, thereby best explaining the red fraction variation. Optimizing for this gives \( \log \Lambda = \log \Sigma + \alpha \log CC + \beta \log AGE\) with \(\alpha = 0.48, \beta = 0.09\). See section \ref{ssec:methodology_vmax}  for meaning of the term \(1/V_{max}\) corrected. }   
	\label{fig:discussion_hist_lambda}
\end{figure}
\begin{figure}
	\centering
	\includegraphics[width=0.5\textwidth]{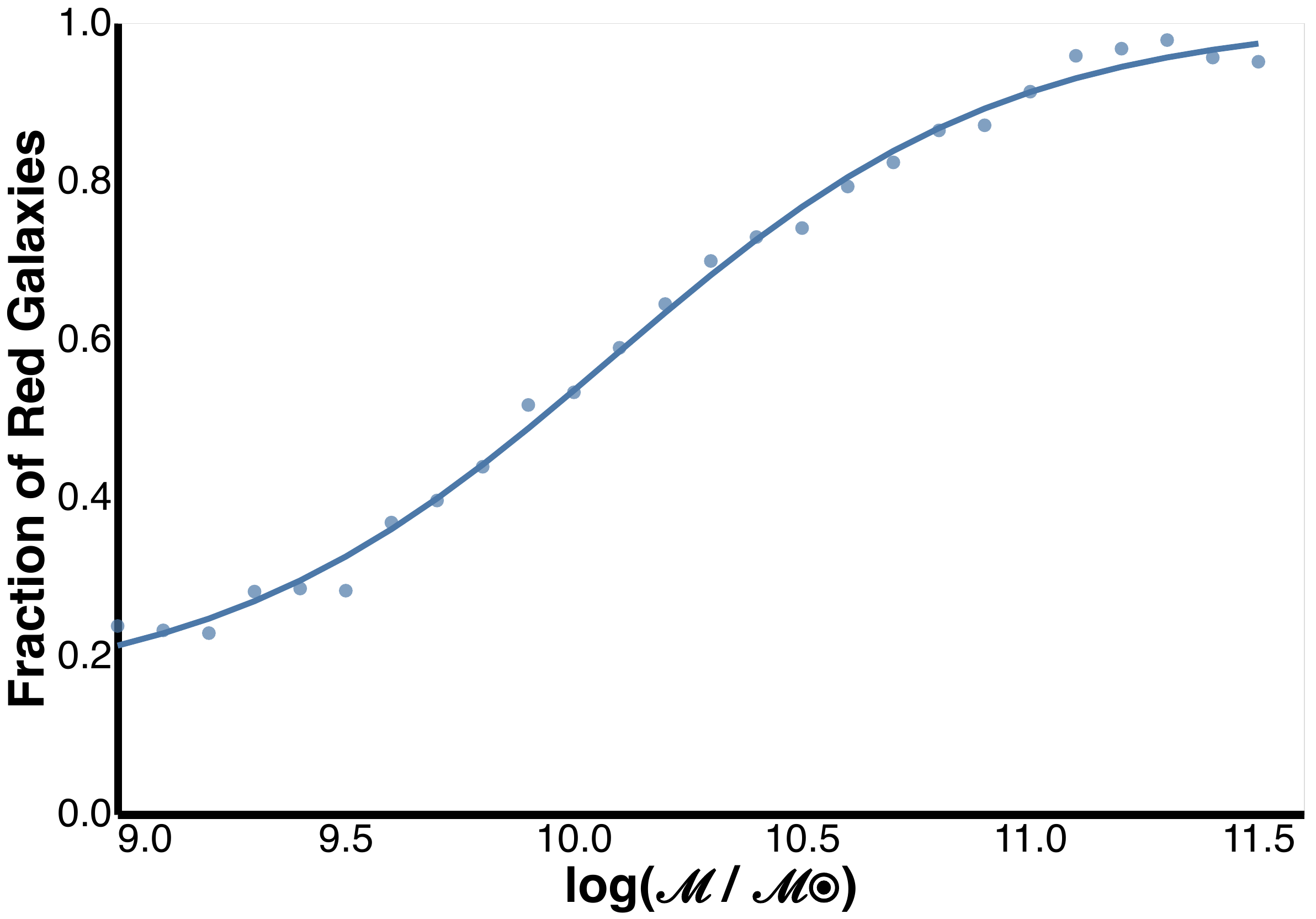}
	\caption{The dependence of the red galaxy fraction on the stellar mass \(\log (\mathcal{M}/\mathcal{M_\odot})\), ignoring (or averaging) the effects of local and geometric environments. The points are the estimations for different mass bins, while the curve is the best fit obtained using the generalized sigmoid function defined in Eqn \ref{eqn:discussion_generalized_sigmoid}}
	\label{fig:discussion_rf_avg_stellar_mass_sigmoid_fit}
\end{figure}

\begin{figure*}
	\centering
	\includegraphics[width=\textwidth]{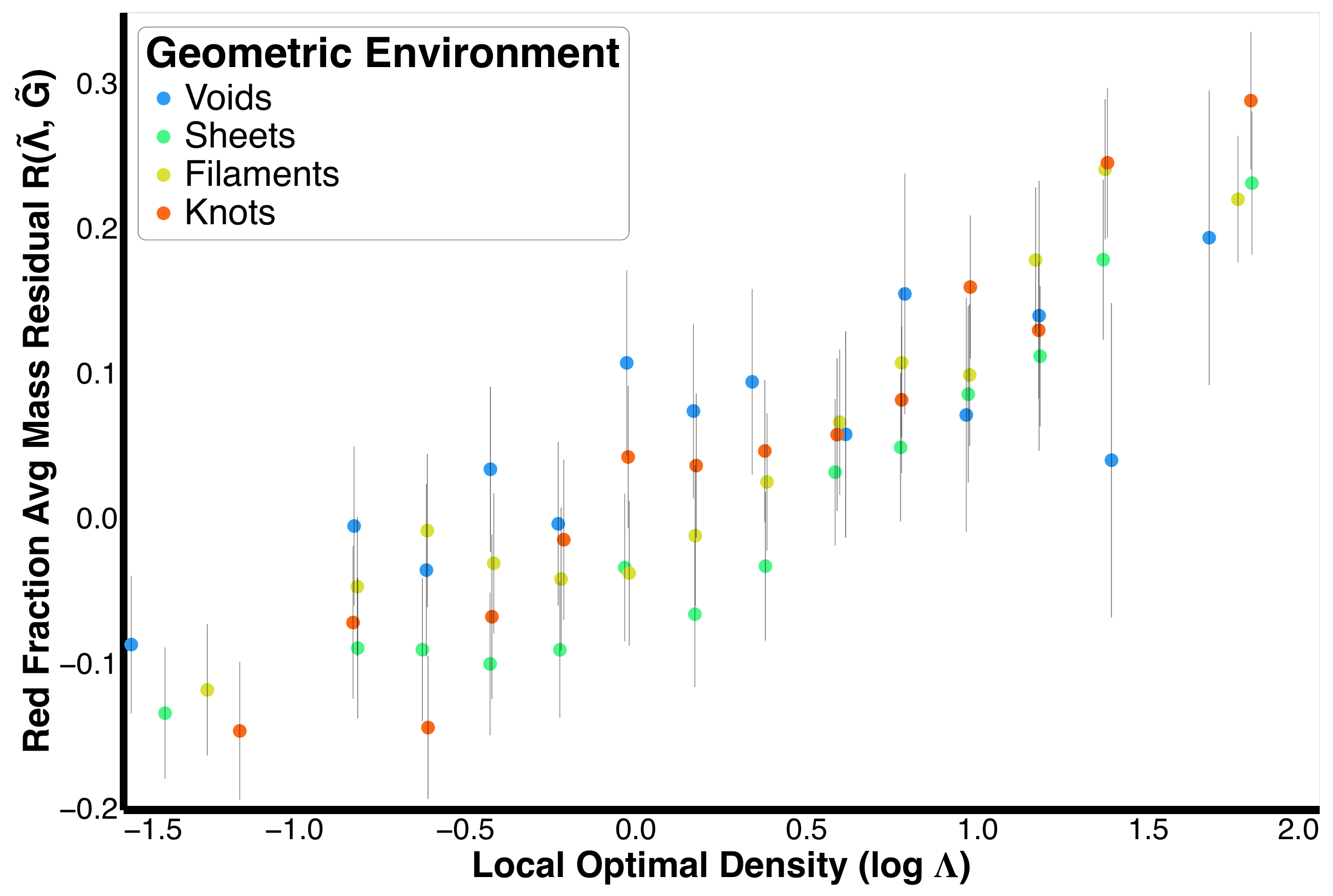}
	\caption{The residual of the red galaxy fraction after subtracting the average stellar mass contribution (based on the fit in Figure \ref{fig:discussion_rf_avg_stellar_mass_sigmoid_fit}). The residual is plotted as a function of the local optimal environment \(\log \Lambda\) (shown on x-axis) and the geometric environment (shown by various colours). The plot suggests that geometric environment does have an effect on the red fraction. In particular, at low to medium local densities, voids tend to have a larger fraction of their galaxies as red than other geometric environments. The error bars are 1-sigma standard deviation errors obtained using jackknife resampling and incorporate the effect of sample size within each environment bin.
	}   
	\label{fig:discussion_punchline_figure}
\end{figure*}

In this section, we discuss the results of our investigation. For all plots, we bin the mass and environment variables (except the discrete-valued geometric environment) to increase the signal-to-noise (S/N) ratio of our estimates. Additionally, we omit points (or grids in the heatmaps) that represent fewer than 20 galaxies, to ensure that statistically significant conclusions can be drawn from the results. 

\begin{table}
	\centering
	\setlength\doublerulesep{0.5pt}
	\def\arraystretch{1.5}
	\begin{tabular}{|c|c|}\hline
		\textbf{Environmental Density} & \makecell{\textbf{Red Fraction range} \\(\textbf{averaged across stellar mass)}} \\\hline
		5th Nearest Neighbour (\(\Sigma\)) & 0.492 \(\pm\) 0.007 
		\\\hline
		Cylindrical Count (CC) & 0.499 \(\pm\) 0.007 
		\\\hline
		Adaptive Gaussian Ellipsoid (AGE) & 0.408 \(\pm\) 0.008 
		\\\hline
		Geometric Environment GeoS10 & 0.150 \(\pm\) 0.007 
		\\\hhline{==}
		\(\Sigma\) vs CC & 0.500 \(\pm\) 0.007 
		\\\hline
		\(\Sigma\) vs AGE & 0.509 \(\pm\) 0.010 
		\\\hline
		CC vs AGE & 0.475 \(\pm\) 0.011 
		\\\hhline{==}
		Local Optimal Density (\(\Lambda\)) & 0.521 \(\pm\) 0.007 
		\\\hline
	\end{tabular}
	\caption{Red Fraction ranges (difference between red fractions at highest and lowest density bins) for different individual environmental measures. The uncertainty estimates are obtained through jackknife resampling. Also shown are ranges when contrasting two different measures of environment. The local optimal density (\(\Lambda\)) has the highest red fraction range. 
	}
	\label{tab:discussion_red_fraction_range} 
\end{table}

\subsection{Effect of environment on galaxy colour}
\label{ssec:discussion_env_effect}

Figure \ref{fig:discussion_1d_plots} shows the variation of galaxy colour with the stellar mass and the environmental densities. As previously mentioned, we use the fraction of galaxies that are red within a given mass-environment bin as our metric for assessing the colour dependence. 
The general trend from looking at these plots is that both the stellar mass and the environmental density play a key role in explaining the evolution in the red fraction, independent of each other. When looking at a given stellar mass bin, denser regions are dominated by red galaxies. Similarly when looking at regions of a given environmental density, the more massive galaxies are likely to be red.

It is worth asking at this point whether the results in Figure \ref{fig:discussion_1d_plots} translate from galaxy colour to galaxy star formation rates (SFR). This could be interesting because observational SFR indicators may probe different intrinsic properties of galaxies. We used a specific star formation rate (SSFR) indicator available in the GAMA DR4 \texttt{MagPhys} catalog instead of colour. We repeated our analysis using the methodology described in section \ref{ssec:methodology_separating_populations} to separate the population into quenched/star-forming subpopulations. We estimated the quenched fraction equivalent to the red fraction and studied its dependence on the environment and mass, similar to Figure \ref{fig:discussion_1d_plots}. Our analysis did not yield any new information compared to the results obtained using galaxy colour. This suggests that the SSFR is not necessarrily more reliable at selecting quenched galaxies.

\subsection{Comparing different environments}
\label{ssec:discussion_compare_envs}

Even though there is some variation in red fraction for all environments in Figure \ref{fig:discussion_1d_plots}, the largest dependence occur for the 5th Nearest Neighbour Surface density \(\Sigma\) and the Cylindrical Count (CC). The effect is weaker for the Adaptive Gaussian Ellipsoid (AGE) density, and the weakest for geometric environments. For a given mass bin, the change in red fraction is less stark as we move from voids to knots, as defined by the \citet{eardleyGalaxyMassAssembly2015} method. 

We can do a quantitative check of this by estimating for each density measure the average red fraction range, i.e. the difference in the red fraction between the lowest and highest density regions. Note that we do not consider the stellar mass variation for this calculation. We can do this because all the environment measures in this approach are computed for the same set of objects and have the same stellar mass range. 

The red fraction range is easy to calculate for the geometric measure, which has predefined discrete bins. In contrast, the local environment measures are continuous values, and we must bin the data. To ensure a fair comparison, we choose bins such that the lowest and highest bins contain the same number of objects across the individual environments, and choose the remaining bins to be of the same size of 0.2. For the lowest bin, we set this number to be 25\% of all objects in our sample, and for the highest bin, we set it to be 5\%. The lowest bin fraction is kept high because the Cylindrical Count measure has a value of 0 - its least possible value - for a quarter of its samples, as mentioned in section \ref{sssec:data_cc}.  The resulting red fraction ranges are shown in Table \ref{tab:discussion_red_fraction_range}, with the uncertainty estimates obtained through jackknife resampling. The numbers show that there is varying information across different environmental measures. While the 5th Nearest Neighbour Surface density and the Cylindrical Count density have similar red fraction ranges, the range for the Adaptive Gaussian Ellipsoid density is lower - with \(\sim 8 \sigma\) difference from the previous two densities. 
Table \ref{tab:discussion_red_fraction_range} also shows the red fraction range for the geometric environment, although we not that it is not directly comparable to the range for local measures since the bin sizes are different.

We can also assess the similarity of information provided by each density measure by looking at how the red fraction compares as a function of two different environmental measurements. Figure \ref{fig:discussion_2d_plots} shows this as heatmaps. (Here too we are not studying the dependence across stellar mass). 
Analyzing these plots, we see that there is variation in the red fraction along each environment axis, increasing when moving from lower to higher density regions. Interestingly, if we focus on the heatmaps containing only the local environmental measures (Figures \ref{fig:discussion_chart_5nn_cc}, \ref{fig:discussion_chart_5nn_age} and \ref{fig:discussion_chart_cc_age}), we find that the maximum variation occurs as we move diagonally, from the bottom right (lowest bin for both densities) to the top left (highest bin for both densities). This suggests that all three measures contain different information and are complementing each other. We might therefore be able to define a new environmental measure that combines these three measures and better explains the variation in red fraction. 
The slightly higher red fraction ranges when comparing the local densities (especially the entries corresponding to \(\Sigma\) vs CC and \(\Sigma\) vs AGE) as shown in Table \ref{tab:discussion_red_fraction_range} also support this, although we note that the number of objects in the lowest and highest bins vary by a few percent for these results. 

By contrast, if we focus on the heatmaps involving the geometric environment (Figures \ref{fig:discussion_chart_5nn_geos10}, \ref{fig:discussion_chart_cc_geos10}, \ref{fig:discussion_chart_age_geos10}), we see that the variation is lower along the axis of geometric environment. This suggests that the red fraction is relatively less dependent on geometric environment than on local measures.
 

\subsection{Testing the effect of geometric environment}
\label{ssec:discussion_test_effect_geom_env}


We have seen that the effect on galaxy colour is more pronounced for local environment densities (Figures \ref{fig:discussion_chart_5nn}, \ref{fig:discussion_chart_cc} and \ref{fig:discussion_chart_age}) than the geometric environment (Figure \ref{fig:discussion_chart_geos10}). Here we probe this further by asking whether the geometric environment adds any significant information beyond what is available in the local environment. Testing for this is important for galaxy evolution models. The cosmic web affects galaxy stellar mass (e.g. \citealt{kauffmannEnvironmentalDependenceRelations2004}) and morphological properties (e.g. \citealt{dasDetectionMolecularGas2015}). It is therefore not unreasonable to expect that large scale flows might influence galaxy evolution. If they do, this information is most likely to be contained in the geometric environment (e.g. \citealt{wetzelClusteringMassiveHalos2007,wetzelClosePairsProxies2008,welkerSAMIGalaxySurvey2020}). There is previous work done to identify the effect of large scale environment on dark matter halos (e.g. \citealt{hellwingCaughtCosmicWeb2021}), suggesting there is interest in the question we seek to answer here. 

We define \(\Lambda\) to be the "optimal" local environment measure - the maximal information we can extract out of local environment. Given that the three local environmental measures seem to contain somewhat different information, we model \(\Lambda\) as a linear combination of these (in log space)

\begin{equation}
    \log \Lambda = \log \Sigma + \alpha \log CC + \beta \log AGE
\label{eqn:discussion_local_optimal_density}
\end{equation}
Note that since our aim is to maximise the red fraction range, the two parameters in equation \ref{eqn:discussion_local_optimal_density} are sufficient for optimizing - the coefficient on \(\log \Sigma\) can be set to 1 without loss of generality.

Quantitatively, optimal density means one which best explains the variation in the red fraction - specifically whose distribution maximizes the red fraction range. We bin the optimal density in the same way we binned the individual densities. To seek the optimal parameters we first did a coarse search of \(10^{-3} < \alpha, \beta < 10\), which determined that the optimal parameters lay in the ranges \(0.1 < \alpha < 10, 0.01 < \beta < 1\). A subsequent fine grained search of the region yielded best results for \(\alpha = 0.48\) and \(\beta = 0.09\). For the coarse and fine searches, we found the red fraction range to be continuous in \(\alpha\) and \(\beta\), with only a single notable minima. It makes sense that the optimal coefficient for the cylindrical count is close to 1 (at an order of magnitude level) since it seems to do an equally good job as the nearest neighbour density in explaining the red fraction variation. 

Figure \ref{fig:discussion_hist_lambda} shows the histogram of the local optimal density \( \Lambda\). From Table \ref{tab:discussion_red_fraction_range}, we see that the maximal red fraction range obtained for \(\Lambda\) is
\(0.521 \pm 0.007\). This is \(2.2\sigma\) higher than the range obtained from the Cylindrical Count density and \(2.9\sigma\) higher than the range obtained from the 5th Nearest Neighbour Surface density. We therefore recommend using \(\Lambda\) in applications involving precision measurements of the effect of local environment on the red galaxy fraction, e.g. if one aims to remove the local environmental dependence. 

In the discussion below, \(\tilde{\Lambda}, \tilde{G}, \tilde{\mathcal{M}}\) stand for bins of the local environment \((\log \Lambda)\), geometric environment and stellar mass \((\log \frac{\mathcal{M}}{\mathcal{M_\odot}})\) respectively.\,\(r(\boldsymbol{\theta})\) is the red fraction estimate as a function of one or more parameters \(\boldsymbol{\theta}\). 

We can now look for the effect of geometric environment beyond that accounted for by the local environment and the stellar mass. The environmental density comparisons in Figure \ref{fig:discussion_2d_plots} show that the median (mode in case of geometric environment) environmental density varies with mass, so we need to control for this effect. We do that by removing the red fraction dependence on stellar mass and examine how the residuals vary with the environment, i.e \(r(\tilde{\Lambda}, \tilde{G})\). In order to estimate the stellar mass dependence, we recompute the red fraction values for different stellar mass bins, ignoring the environmental differences 
, i.e. \(r(\tilde{\mathcal{M}})\). The points in Figure \ref{fig:discussion_rf_avg_stellar_mass_sigmoid_fit} show the recomputed red fraction as a function of stellar mass. We found a linear or logistic regression model did not give a good fit to these values, instead we found a generalized sigmoid function to work better (minimized the root mean-squared error) 

\begin{equation}
    r(\tilde{\mathcal{M}}; a, b, c, d) = \frac{a}{1 + \exp({-b \tilde{\mathcal{M}} + c})} + d
    \label{eqn:discussion_generalized_sigmoid}
\end{equation}

The optimal value of parameters obtained was \(a = 0.85, b = 2.29, c =  23.15, d = 0.14\). The corresponding curve is shown as the solid line in Figure \ref{fig:discussion_rf_avg_stellar_mass_sigmoid_fit}.

To remove the effect of mass from the red fraction estimates, we compute \(r(\tilde{\Lambda}, \tilde{G})\) and subtract the average contribution from stellar mass \(\langle r(\tilde{\Lambda}, \tilde{G}) \rangle_{\tilde{\mathcal{M}}}\). We obtain the latter by averaging \(r(\tilde{\mathcal{M}})\) for mass bins of size 0.1, weighted by the number of objects in each \((\tilde{\Lambda}, \tilde{G}, \tilde{\mathcal{M}})\) bin. We label the residuals \(R(\tilde{\Lambda}, \tilde{G})\).

\begin{align}
    R(\tilde{\Lambda}, \tilde{G}) &\equiv r(\tilde{\Lambda}, \tilde{G}) - \langle r(\tilde{\Lambda}, \tilde{G}) \rangle_{\tilde{\mathcal{M}}} \\\nonumber
    &= r(\tilde{\Lambda}, \tilde{G}) - \frac{1}{N} \sum_i r(\tilde{\mathcal{M}}_i) n_i
\end{align}
where \(n_i\) is the number of objects in bin \((\tilde{\Lambda}, \tilde{G}, \tilde{\mathcal{M}}_i)\) and \(N = \sum_i n_i\). 


Figure \ref{fig:discussion_punchline_figure} shows  \(R(\tilde{\Lambda}, \tilde{G})\) as a function of the local environment (x-axis) and the geometric environment (colour). 
The error bars shown in the plot are 1-sigma standard error estimates obtained by jackknife resampling. The jackknife resampling is done both when averaging across \(r(\tilde{\mathcal{M}})\) and also when computing the red fraction estimates \(r(\tilde{\Lambda}, \tilde{G})\). The latter incorporates the uncertainty due to varying sample size across different environment bins. For instance, voids at lowest local densities have \(\sim 3000\) objects, while those at the highest densities have \(\sim 30\) objects. The error bars at higher densities are therefore higher.

The figure suggests a residual geometric environment effect between the different bins, e.g. between voids and sheets. We perform an Analysis of Variance (ANOVA) test to assert whether there is a difference between \(R(\tilde{\Lambda}, \tilde{G})\) for the different geometric environment bins when looking at a fixed local environment bin. The results are shown in Table \ref{tab:anova_results} and confirm the presence of a residual difference at significane level \(\alpha = 0.1\).
The above-average red fraction residuals for voids could be a consequence of the definition of voids as per \cite{eardleyGalaxyMassAssembly2015} and may require further investigation (see \ref{ssec:future_alternate_geom_envs}).
Conversely, the result may indicate a potential effect of the surrounding large scale structure - perhaps a mechanism for replenishing star-forming gas, whose absence in voids leads to larger fraction of red galaxies. 
At higher local densities the trend is reversed - voids have lower residuals than other geometric bins, though the error bars are also wider. 
Finally, the residuals increase as we go from lower to higher local density, reaffirming a positive correlation of red fraction with local environment independent of mass.

\begin{table*}
	\centering
	\setlength\doublerulesep{0.5pt}
	\def\arraystretch{1.5}
	\begin{tabular}{|c|c|c|c|c|c|c|c|c|}
		\hline
		\(\boldsymbol{\log \Lambda} \) &       \makecell{\textbf{Sum of Squares} \\\textbf{within group}} &       \makecell{\textbf{Mean of Squares} \\\textbf{within group}}  &        \makecell{\textbf{Sum of Squares} \\\textbf{between groups}}  &       \makecell{\textbf{Mean of Squares} \\\textbf{between groups}}  &            \textbf{F-statistic}&  \textbf{\(\text{df}_1\)} &           \textbf{\(\text{df}_2\)} &  \makecell{\textbf{Significant at}\\ \(\boldsymbol{\alpha=0.1}\)?} \\
		\hline
	  -2.746 &  38.757 &  0.002 &   6.640 &  2.213 &  1048.209 &    3 &  18354.465 &                       Yes \\\hline
		 -0.919 &  11.689 &  0.002 &   3.755 &  1.251 &   503.467 &    3 &   4701.871 &                       Yes \\\hline
	 -0.719 &  13.116 &  0.002 &  12.948 &  4.316 &  1584.159 &    3 &   4814.153 &                       Yes \\\hline
	 -0.519 &  14.234&  0.002 &   9.468 &  3.156 &  1209.381 &    3 &   5454.602 &                       Yes \\\hline
		  -0.319 &  14.253 &  0.002 &   5.073 &  1.691 &   656.601 &    3 &   5534.304 &                       Yes \\\hline
		  -0.119 &  13.631 &  0.002&  14.616 &  4.872 &  1869.357 &    3 &   5229.968 &                       Yes \\\hline
	  0.081 &  11.569 &  0.002 &   8.267 &  2.755 &  1099.752 &    3 &   4616.798 &                       Yes \\\hline
		 0.281 &   9.795 &  0.002 &   4.258 &  1.419 &   575.594 &    3 &   3971.571 &                       Yes \\\hline
		  0.481 &   9.135 &  0.002 &   0.535 &  0.178 &    65.414 &    3 &   3350.898 &                       Yes \\\hline
		  0.681 &  10.407 &  0.002 &   2.392 &  0.797 &   289.701 &    3 &   3780.694 &                       Yes \\\hline
	   0.881 &   8.687 &  0.002 &   3.707 &  1.235 &   448.259 &    3 &   3151.066 &                       Yes \\\hline
	   1.081 &   6.806 &  0.002 &   2.355 &  0.785 &   313.924 &    3 &   2721.003 &                       Yes \\\hline
		   1.281 &   6.459 &  0.002 &   9.644 &  3.214 &  1186.889 &    3 &   2385.002 &                       Yes \\\hline
		    1.481 &  10.863 &  0.002 &   7.507 &  2.502 &  1145.862 &    3 &   4974.112 &                       Yes \\
		\hline
	\end{tabular}
	\caption{Analysis of Variance (ANOVA) result to test whether there is a difference in \(R(\tilde{\Lambda}, \tilde{G})\) for the different geometric environment bins when looking at a fixed local environment \(\log \Lambda\). The critical F-statistic for \(\text{df}_1 = 3, \text{df}_2 > 120\) at significance level \(\alpha = 0.1\) is 3.78. There is therefore a significant residual difference at each local environment bin. 
	}
	\label{tab:anova_results} 
\end{table*}

\section{Conclusions and Future Work}
\label{future_work_conclusion}

In this work, we studied the effect of environment on the restframe galaxy colour, using the fraction of red galaxies as a metric. We found that the red fraction varies with the local environment independently of the stellar mass. For a given stellar mass, galaxies in a high density environment are more likely to be red when compared to those in a low density environment. Possible explanations include loss of star forming gas due to stripping in denser environments, through mechanisms such as strangulation, galaxy harassment and ram pressure stripping. We also found that the different environmental measures defined in GAMA contain different information. In particular, the local environmental measures exhibit more red fraction variation as we move from regions of lower to higher densities, when compared to the geometric environment. However, when the different local measures are combined to produce an "optimal" local density, there is a residual effect of the geometric environment independent of the local environment and stellar mass. 

\subsection{Future Work}


The following are potential directions for further research:

\subsubsection{Investigate the effect of alternate definitions of large scale environment }
\label{ssec:future_alternate_geom_envs}
The higher value of \(R(\tilde{\Lambda}, \tilde{G})\) for voids at a given local density is dependent on the definition of voids within the geometric environment framework of \cite{eardleyGalaxyMassAssembly2015}, i.e. voids being regions surrounded by similar or higher density regions in all directions. It would be useful to repeat our analysis of section \ref{ssec:discussion_test_effect_geom_env} using alternative means of defining large scale environment, in order to determine whether the apparent strong influence of voids is robust to variations in methodology.

\subsubsection{Studying the residual effect of dark matter haloes on galaxy colour}
\label{ssec:future_dark_haloes}




The red galaxy fraction is also expected to be affected by dark matter halo mass, with more massive halos expected to have more red galaxies (see e.g. \citealt{behrooziUniverseMachineCorrelationGalaxy2019}). The environmental density certainly plays a role in this relation - more massive dark matter halos would form out of bigger overdensities 
, which would accrue more baryons, forming more galaxies and thus in general have higher environmental densities. For instance, \cite{baldryGalaxyBimodalityStellar2006} found a correlation between the overdensity 
and the projected 5th Nearest Neighbour environmental density \(\Sigma\) as estimated from simulations. 

It would therefore be interesting to study whether the halo mass has any residual correlation with the red galaxy fraction, after accounting for the effects of environment and stellar mass. If a correlation is found, it may suggest that massive dark matter haloes have some inherent mechanism that affects the evolution of galaxies and makes them more likely to be quenched. 


\subsubsection{Studying the environmental effects via simulations}
\label{ssec:future_simulations}

Future work could also look into seeing how well simulations explain the observed environmental effects. In particular, we would look to see whether simulations also show a small residual dependence on large scale geometric environment, and if so what mechanism is causing this. This would be similar to the approach of \cite{baldryGalaxyBimodalityStellar2006}, but for all the environmental densities used in this work. For e.g., the geometric environmental measure is a conceptually different metric, so it would be intriguing to see how well it is reproduced in analytical models, especially given that it has a small effect on the red fraction. Any environmental metric which is measurable in observational data can be reproduced in a cosmological simulation, which is an important step in any robust comparison with simulations. 
In this case, it would involve identifying mechanisms that explain the observed red fraction trends found in the different environmental measures. It would also involve comparing the different galaxy evolution models with their different feedback mechanisms (e.g. \citealt{bowerBreakingHierarchyGalaxy2006,2008MNRAS.391..481S,vandenboschImportanceSatelliteQuenching2008,pengStrangulationPrimaryMechanism2015,bianconiStarFormationBlack2016,2017MNRAS.465...32B}) to see which models more closely align with the results shown here.

\section*{Acknowledgements}
GAMA is a joint European-Australasian project based around a spectroscopic campaign using the Anglo-Australian Telescope. The GAMA input catalogue is based on data taken from the Sloan Digital Sky Survey and the UKIRT Infrared Deep Sky Survey. Complementary imaging of the GAMA regions is being obtained by a number of independent survey programmes including GALEX MIS, VST KiDS, VISTA VIKING, WISE, Herschel-ATLAS, GMRT and ASKAP providing UV to radio coverage. GAMA is funded by the STFC (UK), the ARC (Australia), the AAO, and the participating institutions. The GAMA website is http://www.gama-survey.org/ . 

Our work builds on many software tools/libraries, which we acknowledge by citing. Our analysis was done in the Python ecosystem, with the following packages being the most relevant:

\begin{itemize}
	\item For the environment: Anaconda \citep{anaconda}, Jupyter Notebook \citep{Kluyver2016jupyter}
	
	\item For the analysis: Numpy \citep{harrisArrayProgrammingNumPy2020}, Scipy \citep{virtanenSciPyFundamentalAlgorithms2020}, Pandas \citep{mckinneyDataStructuresStatistical2010}, Astropy \citep{2018AJ....156..123A,2013A&A...558A..33A}, Scikit-learn \citep{JMLR:v12:pedregosa11a}, Stan \citep{pystan,stanStanModelingLanguage2021}
	
	\item For plotting and visualization: Altair \citep{VanderPlas2018}, Seaborn \citep{Waskom2021}, Matplotlib \citep{hunterMatplotlib2DGraphics2007}, Plotly \citep{plotly} 
\end{itemize}

\section*{Data Availability}

This work uses publicly available data as part of data release 4 (DR4) of the GAMA survey. DR4 can be accessed at http://www.gama-survey.org/dr4/ 
 



\bibliographystyle{mnras}
\bibliography{astro_paper} 




\appendix

\section{How the red fraction estimates vary among the different colours}
\label{ssec:discussion_effect_colour}

\begin{figure}
	\centering
		\begin{subfigure}[b]{\columnwidth}   
			\centering 
			\includegraphics[width=\textwidth]{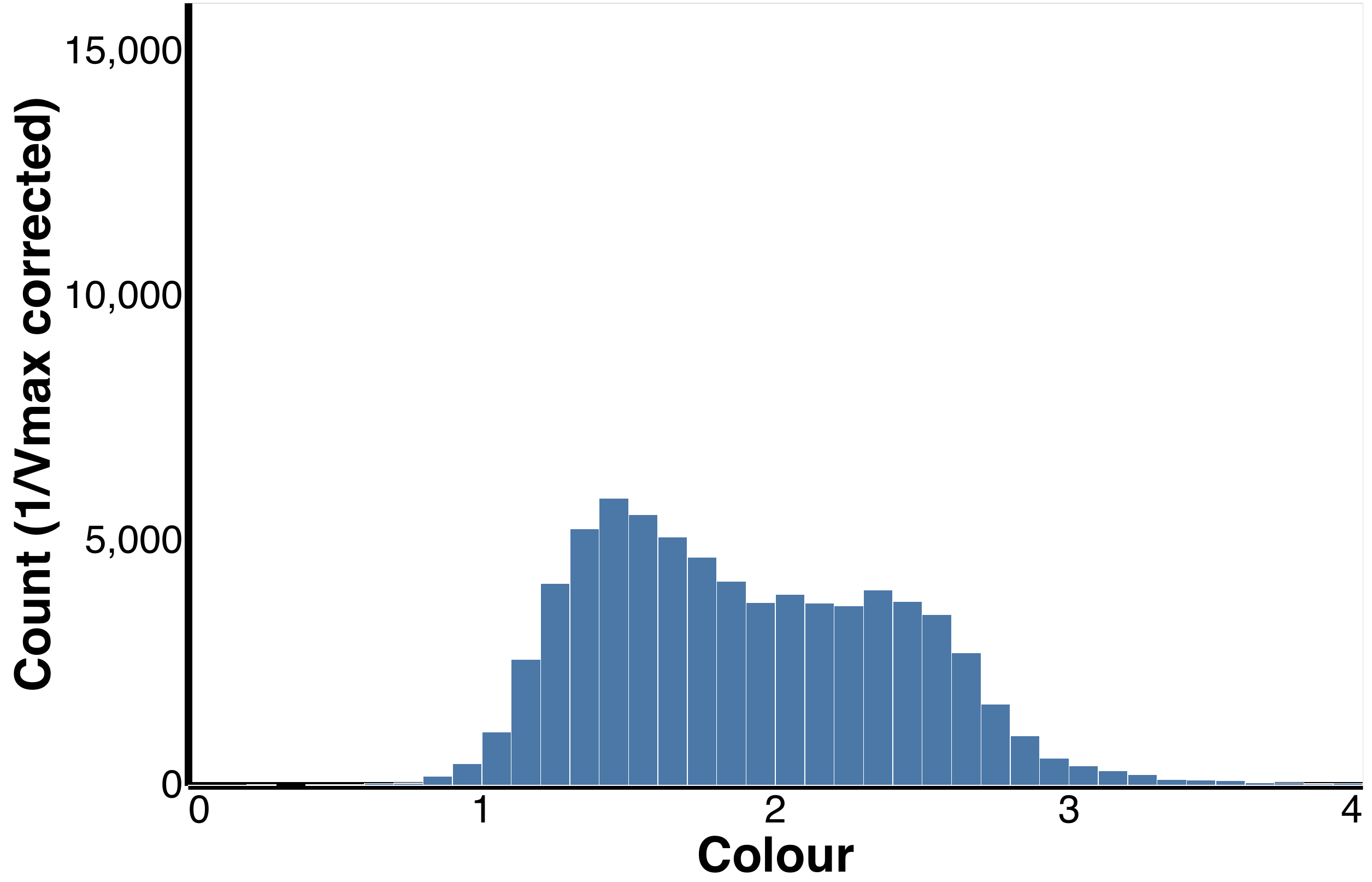}
			\caption[]%
			{{\small \(u_{model} - r_{model}\)   }}    
			\label{fig:appendix_hist_uminusr_model}
		\end{subfigure}
		\vskip\baselineskip
		\begin{subfigure}[b]{\columnwidth}   
			\centering 
			\includegraphics[width=\textwidth]{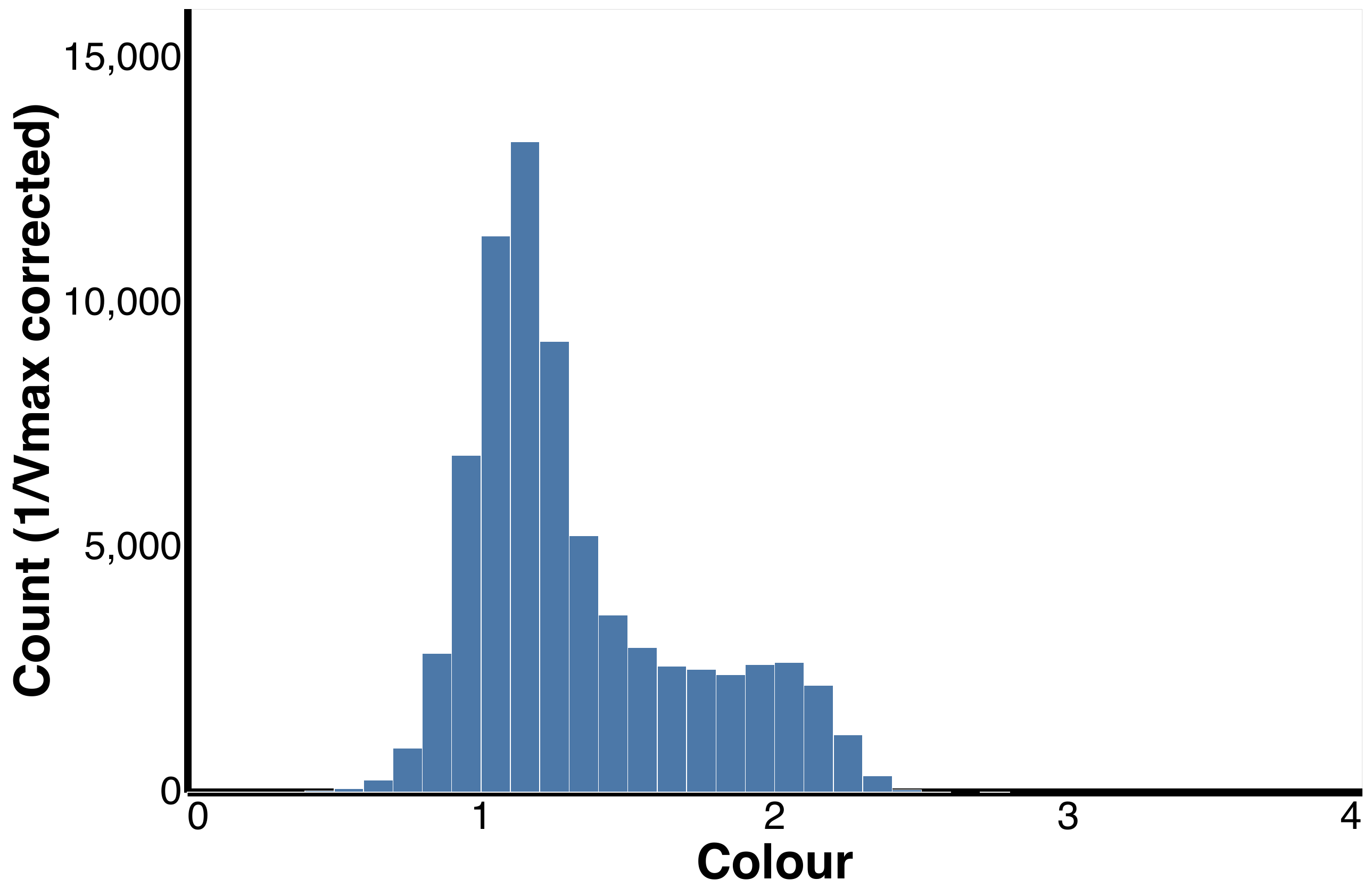}
			\caption[]%
			{{\small \(u^* - r^*\) }}    
			\label{fig:appendix_hist_uminusr_stars}
		\end{subfigure}
	\vskip\baselineskip
		\begin{subfigure}[b]{\columnwidth}   
			\centering 
			\includegraphics[width=\textwidth]{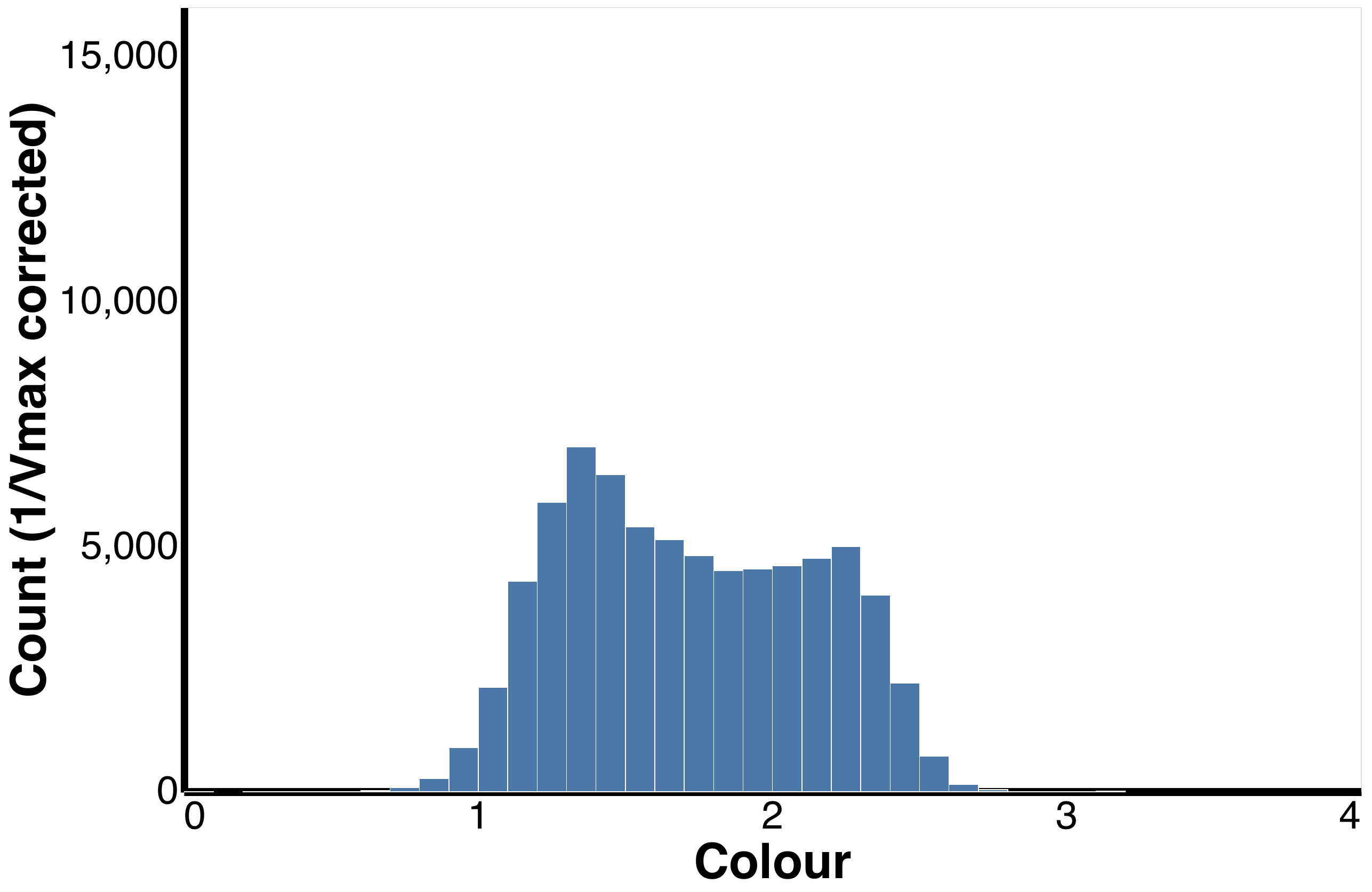}
			\caption[]%
			{{\small \(u - r\)}}    
			\label{fig:appendix_hist_uminusr}
		\end{subfigure}
	\caption[ ]
	{{ \small Histograms of the different colours used to study the red fraction dependence - (a) \(u_{model} - r_{model}\), (b) \(u^* - r^*\) and (c) \(u - r\). The counts shown here are volume (\(1/V_{max}\)) weighted, see section \ref{ssec:methodology_vmax} for more information on how this correction is done. }}
	\label{fig:appendix_hist_colour}
\end{figure}

\begin{figure}
	\centering
		\begin{subfigure}[b]{\columnwidth}   
			\centering 
			\includegraphics[width=\textwidth]{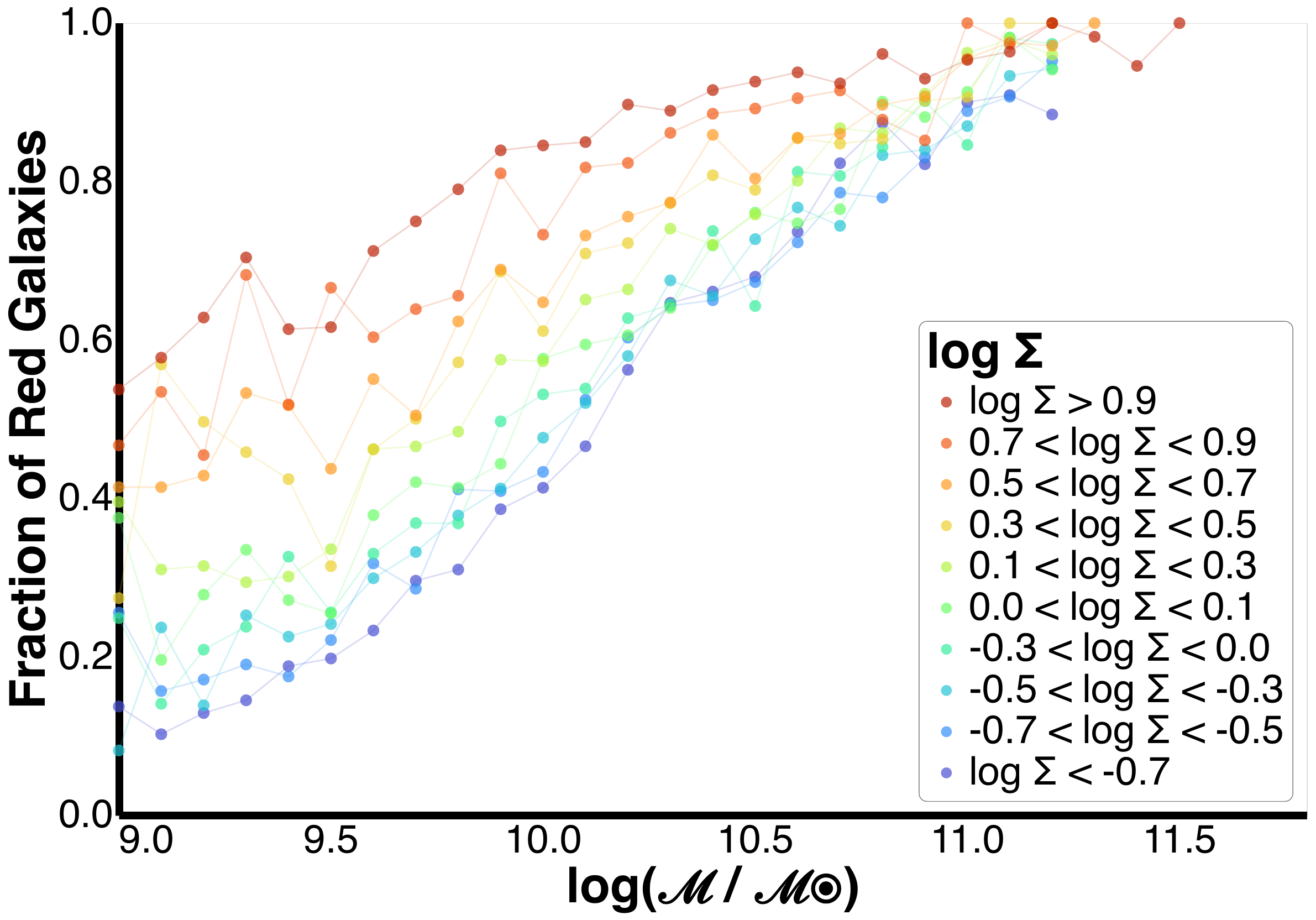}
			\caption[]%
			{{\small \(u-r\)}}    
			\label{fig:discussion_chart_uminusr}
		\end{subfigure}
		\vskip\baselineskip
		\begin{subfigure}[b]{\columnwidth}   
			\centering 
			\includegraphics[width=\textwidth]{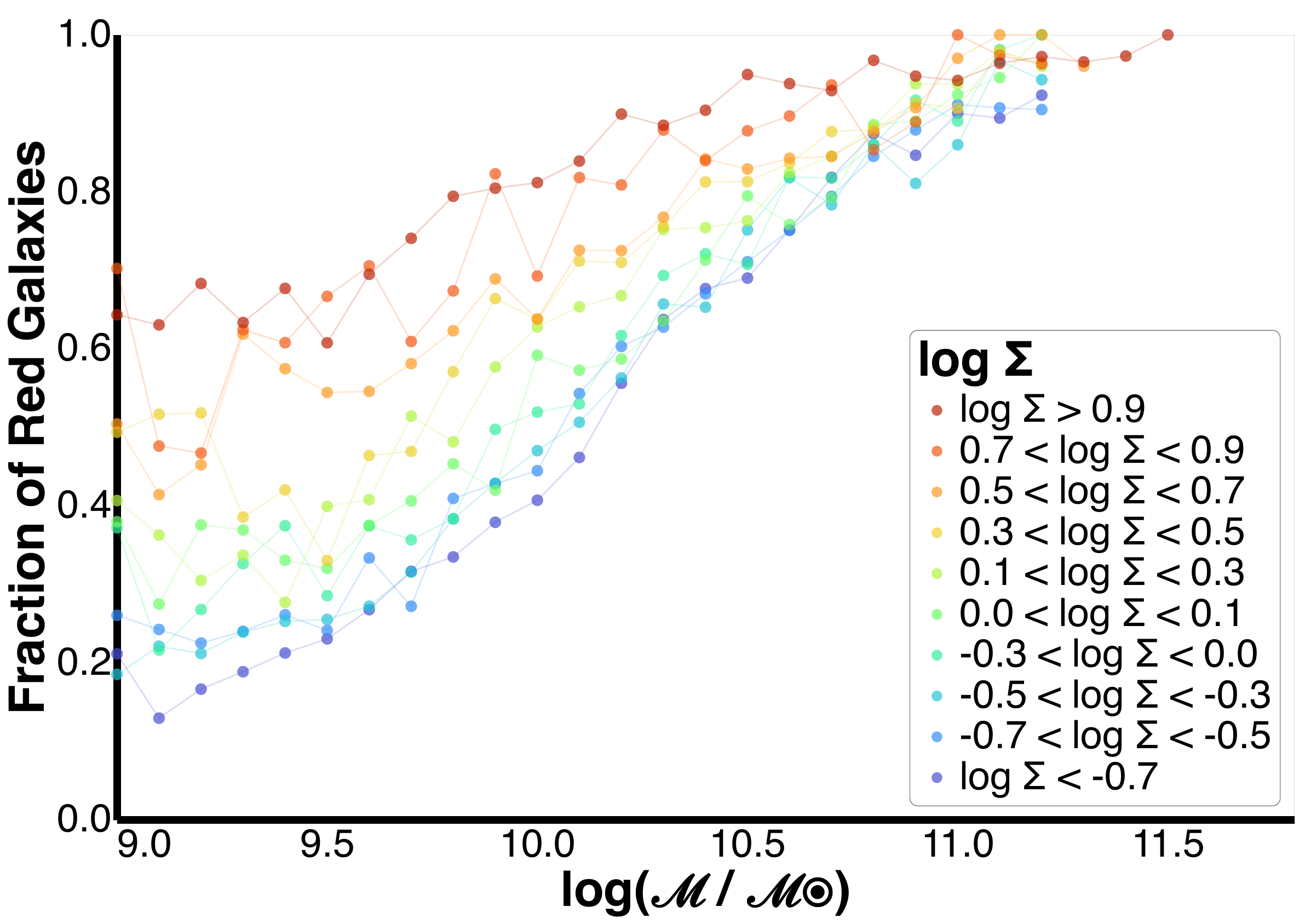}
			\caption[]%
			{{\small \(u_{model} - r_{model}\)}}    
			\label{fig:discussion_chart_uminusr_model}
		\end{subfigure}
		\vskip\baselineskip
		\begin{subfigure}[b]{\columnwidth}   
			\centering 
			\includegraphics[width=\textwidth]{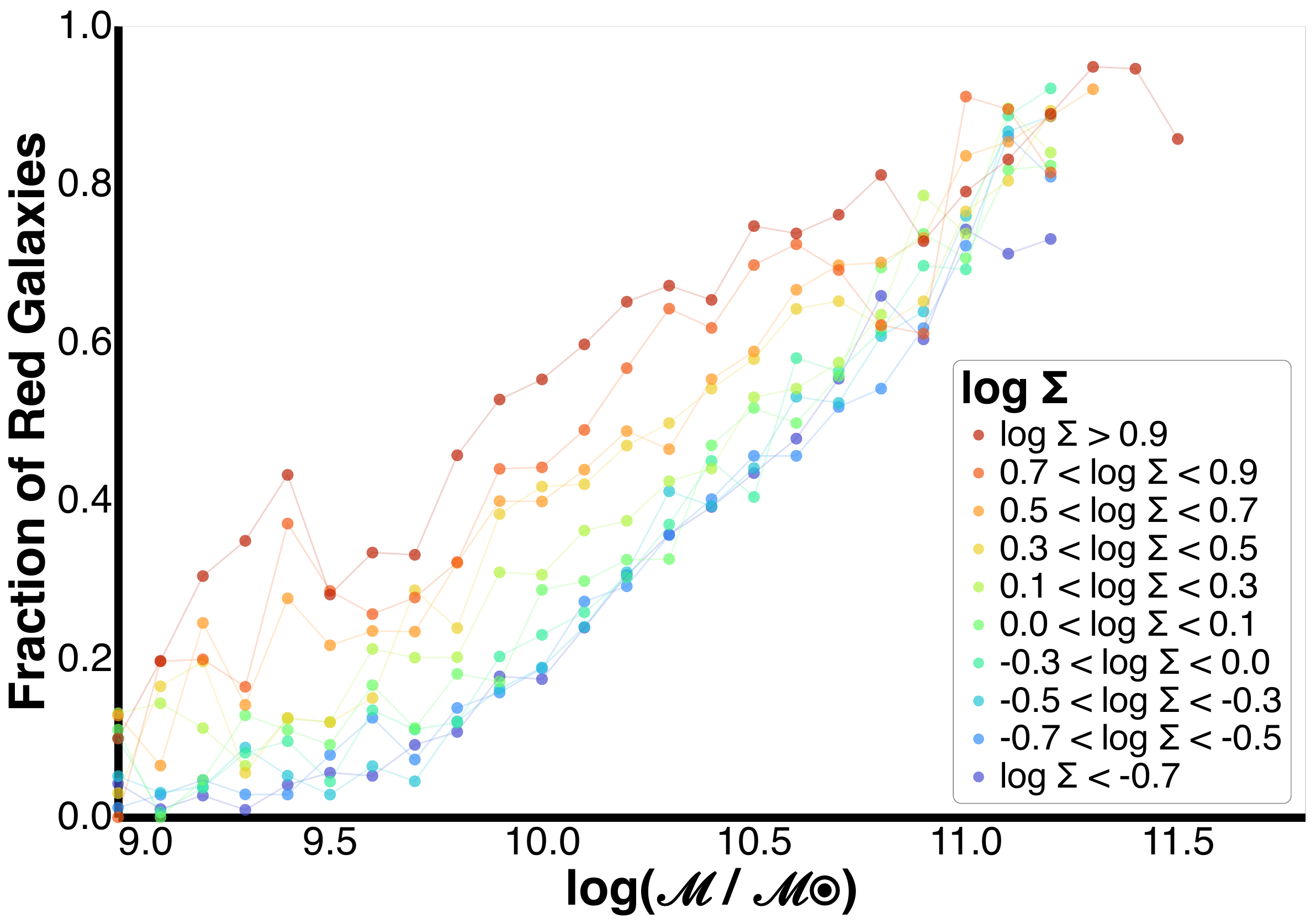}
			\caption[]%
			{{\small \(u^* - r^*\) }}    
			\label{fig:discussion_chart_uminusr_stars}
		\end{subfigure}
		\hfill
	\caption[ ]
	{{ \small Effect of using different colours on the fraction of red galaxies as a function of stellar mass \(\log (\mathcal{M}/\mathcal{M_\odot})\) and 5th Nearest Neighbour density (\(\Sigma\)). The labels are similar to those shown in Figure \ref{fig:discussion_1d_plots}. Using \(u^* - r^*\) results in low red fraction estimates in high environmental regions, especially at low mass. This is contrary to what is known in the literature \citep{baldryGalaxyBimodalityStellar2006} and warrants further investigation into how the colour estimates are generated.}}
	\label{fig:discussion_color_effect}
\end{figure}




In our analysis above, we used the colour \(u - r\) which is obtained from LAMBDAR photometry, k-corrected as well as corrected for Milky-Way dust extinction. We also experimented with two additional colours and studied their dependence on the red galaxy fraction. These two colours are defined below:

\begin{itemize}
    \item \textbf{Based on SDSS Model magnitudes} \(\boldsymbol{u_{model} - r_{model}}\): 
This colour is estimated from model magnitudes provided as part of GAMA from DR7 of the Sloan Digital Sky Survey (SDSS). These are corrected for Milky Way dust extinction. We also perform k-correction to redshift \(z = 0\). 
Figure \ref{fig:appendix_hist_uminusr_model} shows a histogram of the resulting colour values. This colour term is similar to that used by \cite{baldryGalaxyBimodalityStellar2006}. 



\item \textbf{Based on GAMA magnitudes with host galaxy dust extinction correction \(\boldsymbol{u^* - r^*}\)}:
The estimates for this colour are obtained similar to \(u - r\), but then are additionally corrected for the light absorbed by dust in the host galaxy. This correction is done using stellar population synthesis (SPS) models \citep{taylorGalaxyMassAssembly2011}. In this type of modeling, there can be degeneracy between internal dust reddening and age/metallicity effects. Figure \ref{fig:appendix_hist_uminusr_stars} shows a histogram of the resulting colour values. See \cite{taylorGalaxyMassAssembly2015} for more details.
\end{itemize}

The histogram of \(u - r\) is shown again in Figure \ref{fig:appendix_hist_uminusr} for comparison. We note that for all the colours, the same approach described in section \ref{ssec:methodology_separating_populations} was used to separate the red and blue populations.

Figure \ref{fig:discussion_color_effect} shows the resulting red fraction estimates.  The plots that use the colours \(u - r\) (Figure \ref{fig:discussion_chart_uminusr}) and \(u_{model} - r_{model}\) (Figure \ref{fig:discussion_chart_uminusr_model}) look similar to each other and to the results obtained by \cite{baldryGalaxyBimodalityStellar2006} (specifically Figure 11(b) of that paper). For high environmental density, we observe high values of red fraction even at low masses. This agreement may be seen as a consistency check for the approach used in this work. 

In contrast the plot for the colour \(u^* - r^*\) (Figure \ref{fig:discussion_chart_uminusr_stars}) has low red fraction values at low masses for high environmental densities. This is different from what has been seen in the literature. As mentioned in \ref{sssec:data_colour}, the colour \(u^* - r^*\) is defined similarly to \(u - r\), except it also attempts to correct for the extinction caused by dust in the host galaxy \citep{taylorGalaxyMassAssembly2015}. This correction involves performing stellar population synthesis (SPS) modeling and is considerably more difficult than estimating the Milky Way dust extinction. Additionally, the degeneracy between dust and the age of a stellar population \citep{bellStellarMasstoLightRatios2001} makes it difficult to model and extract only the dust part of a galaxy, hence separate models must be fit for the stars and the dust. It is therefore possible that there are some unresolved degeneracies in the SPS fitting and the dust extraction which leads to incorrect values of \(u^* - r^*\) colours. This therefore warrants further investigation into how the colour estimates are generated. 

The colours \(u_{model} - r_{model}\) and \(u - r\) produce comparable results, however \cite{taylorGalaxyMassAssembly2011} have shown for SDSS that the magnitudes obtained using model based photometry tend to be biased compared to their Petrosian counterparts, and have therefore recommended using the latter magnitudes. The colour \(u - r\) is computed from GAMA magnitudes obtained using LAMBDAR photometry, which uses elliptical apertures rather than circular ones, and is expected to yield even better estimates than Petrosian magnitudes. For this reason, the colour \(u - r\) was chosen for all the analyses in this work.

\bsp	
\label{lastpage}
\end{document}